\documentclass[utf8]{frontiersFPHY}
\usepackage{url,hyperref,lineno,microtype,subcaption}
\usepackage[onehalfspacing]{setspace}

\usepackage[margin=2cm]{geometry}
\usepackage{wrapfig}
\usepackage{graphicx}
\def\firstAuthorLast{Gu\'elin \& Cernicharo} 
\def\Authors{Michel Gu\'elin\,$^{1,*}$ and Jose Cernicharo\,$^{2}$}
\def\Address{$^{1}$IRAM, St Martin d'H\a`eres, France \\
$^{2}$Instituto de F\'isica Fundamental, CSIC. C/Serrano 121, Madrid, Spain}
\def\corrAuthor{IRAM, 300 rue de la Piscine, 38406 St Martin d'H\a`eres, France}

\def\corrEmail{guelin@iram.fr}

\begin{document}
\onecolumn
\firstpage{-1}

\title[Organic Molecules in Space]{Organic Molecules in Interstellar Space: {\bf Latest advances}} 

\author[\firstAuthorLast ]{\Authors} 
\address{} 
\correspondance{} 

\extraAuth{}

\maketitle

\begin{abstract}
{Although first considered as too diluted for the formation of
molecules {\it in-situ} and too harsh an environment for their
survival, the interstellar medium has turned out to host a rich
palette of molecular species: to date, 256 species, not counting
isotopologues, have been identified. The last decade, and more
particularly the last two years, have seen an explosion of new
detections, including those of a number of complex organic species,
which may be dubbed as prebiotic. Organic molecules have been
discovered not just in interstellar clouds from the Solar
neighbourhood, but also throughout the Milky-Way, as well as in nearby
galaxies, or some of the most distant quasars.
 
These discoveries were made possible by the completion of large
sub-millimetre and radio facilities.  Equipped with new generation
receivers, those instruments have provided the orders of magnitude
leap in sensitivity required to detect the vanishingly weak rotational
lines that allowed the molecule identifications.

Last two years, 30 prebiotic molecules have been detected in
TMC-1, a dust-enshrouded gaseous cloud located at 400 light-years from
the Sun in the Taurus constellation. Ten new molecular species,
have been identified in the arm of a spiral galaxy 6 billion light-yr
distant, and 12 molecular species observed in a quasar at 11 billion
light-yr. We present the latest spectral observations of this outlying
quasar and discuss the implications of those detections
in these 3 archetypal sources.  The basic ingredients involved in the
Miller-Urey experiment and related experiments (H$_2$, H$_2$O, CH$_4$,
NH$_3$, CO, H$_2$S,...) appeared early after the formation of the
first galaxies and are widespread throughout the Universe.  The
chemical composition of the gas in distant galaxies seems not much
different from that in the nearby interstellar clouds. It presumably
comprises, like for TMC-1, aromatic rings and complex organic
molecules putative precursors of the RNA nucleobases, except the
lines of such complex species are too weak to be detected that far.}

\end{abstract}

\section{Introduction}

The question of extraterrestrial life arose as soon as Earth was
recognized as one amid a myriad of celestial bodies. At first, alien
life was fancied as a mere mirror of the life on Earth, with living
creatures possibly scaled to the size of their home planets (see {\it L'autre
Monde} by Savinien de Cyrano de Bergerac, printed in 1657, {{\it Cosmotheoros} by Christiaan Huygens, 
in 1698}, 
and {\it Micromegas} by Voltaire, in 1752). Later, advances in science let writers
realize that aliens may not look so friendly ({\it The War of the
Worlds}, H.G. Wells). Finally, it was argued that
extraterrestrial life could be based on silicium
or germanium, instead of carbon, and take surprising forms. The most imaginative along
this line was F. Hoyle who published, sharp 300 years after Cyrano's
book, a novel that invokes an interstellar cloud as a living being:
{\it The Black Cloud}, 1957.  Sir Fred Hoyle, whose grand reputation
stems from his elucidation of the origin of carbon and heavier
elements ({\it B$^2$FH}, 1957), but also from his taste for unorthodox
theories, was a defender of panspermia and a forceful opponent of
Earth-based abiogenesis.

The panspermia hypothesis, first developped in the 19th century,
assumes that life came from outer space in the form of microscopic
organisms present on meteorites or dust particles that somehow
survived their impact on the Earth. In Hoyle's view, the great advantage
of panspermia was that it immensly increases the time, hence the probability, for a
chance assembly of atoms into prebiotic molecules within the lifetime 
of a (in his view) Steady State Universe. 

Subsequent discoveries have shown that such a process has occured for at least some prebiotic building blocks.
The first interstellar  molecules (CH, CN, CH$^+$) were discovered in the optical spectra of nearby stars 
{(Swings \& Rosenfeld 1937, McKellar 1940, Douglas \& Herzberg 1941)}, but it took {30} years
until more complex species (e.g. NH$_3$ {Cheung et al. 1968}) could be found in interstellar 
space. In echo to Hoyle's science-fiction novel, water and
formaldehyde, the first organic molecule, were detected in 1969 and found
ubiquitous in interstellar clouds (Cheung et al. 1969, Snyder et al. 1969). The construction of the large
(36-ft diameter) radio-telescope on Kitt Peak, Arizona, and of the Bell Laboratories first Schottky barrier diode 
receiver fitted for millimetre-wave observations, opened the way to a surge of new discoveries: CO, 
the most abundant interstellar molecule after H$_2$ (Wilson et  al. 1970), hydrogen sulfide H$_2$S, the first alcohol
CH$_3$OH, ethanol,... 
This firework led Patrick Thaddeus, key actor on this scene, to state the Fisher's Scientific Principle: 
{\it If you can't find 
it at the Fisher Chemical store you will not find it in the interstellar gas} (see Thaddeus 2006). This statement, 
however, was soon overturned by William Klemperer and Thaddeus' own NASA team by the discovery of the first 
{\it non-terrestrial} molecular 
species: protonated carbon monoxyde HCO$^+$, protonated nitrogen N$_2$H$^+$, CCH, C$_3$N, and C$_4$H 
(Herbst \& Klemperer 1974, Green, Montgomery and Thaddeus 1974, Tucker et al. 1974, Gu\'elin and Thaddeus 1977, 
Gu\'elin, Green and Thaddeus 1978), all five unstable species not previoulsy observed in the laboratory.
 
To date, 256 molecular
species, not counting isotopologues, are identified in interstellar clouds and
circumstellar shells (see references hereafter). They are mostly acyclic organic molecules or
radicals with C-chain backbones, as well as half a dozen rings with 5 to 10
C-atoms. They include aldehydes, alcohols, acids, amines and
carboxamides, i.e. the main functional groups needed to initiate the
formation of prebiotic molecules and RNA. 

It is worth noting that no amino acid has been found to date in interstellar space, despite
multiple searches for glycine (NH$_2$-CH$_2$-COOH), the simplest of them (see 
Gu\'elin \& Cernicharo 1989, 
Snyder et al. 2005, Jimenez-Serra et al. 2020). On
the other hand, four putative precursors of glycine, methylamine (CH$_3$NH$_2$-- see Holtom et al. 2005), 
formamide (CH$_3$CHO -- Rubin et al. 1971,
Ferus et al. 2018), glycolonitrile (HOCH$_2$CN --Zeng et al. 2019) and aminoacetonitrile, 
(NH$_2$-CH$_2$-CN -- Belloche et al. 2008)  are detected, 
as is cyanomethanimine (HNCHCN -- Rivilla et al. 2019) a HCN dimer and
possible precursor of adenine, H$_5$C$_5$N$_5$, one of the four nucleobases of DNA.

The increasing rate of new detections mainly stems from
advances in spectroscopic and astronomical instrumentation. Progress
has been particularly striking in the {sub-millimetre and radio domains}.  
Those wavelenghts cover the lowest rotational
lines of molecules of astrochemical interest and allow radiation to
freely transit through dust-enshrouded clouds. In Section 2, we
describe the state-of-the-art instrumentation used in current molecule
searches. In the next Sections, 3 and 4, we present recent results based
on extensive spectral surveys of 3 archetypal interstellar sources:
first a cold dark cloud, not unlike Hoyle's {\it Black Cloud}, in the Solar vicinity, 
then the spiral arm of a
galaxy located 7 billion light-years away, finally a remote quasar
at a time when the Universe was only 2.7 billion {light-years old}.

\vspace{1.0cm}
\section{Detectors and telescopes at mm/sub-mm wavelengths}
In the last decade, the field of astrochemistry has seen major
advances triggered by the completion of {new powerful radio telescopes,
such as the Atacama Large sub-Millimetre Array (ALMA) and the Northern
Extended Millimetre Array (NOEMA), coupled to impressive gains in
sensitivity of receivers and in bandwidth. We here briefly review} the
leading facilities and their new equipment that allowed the latest
progress in searching for new molecules in interstellar space.

The gains in detector sensitivity and bandwidth were linked to the
development of low-noise/wide-band cryogenic amplifiers, based on
HEMTs (High Electron Mobility Transistors) that operate around 15
K, and the development of mixers equipped with SIS
(Superconductor-Insulator-Superconductor) junctions, based on photon-assisted 
electron tunneling at temperatures below 4 K.
{State of the art} heterodyne receivers equipped with such devices now
achieve noise temperatures of only few times the quantum limit
($2h\nu/k$), hence close to optimal, and ally increasingly wide tuning
ranges and instantaneous bandwidths.

The heterodyne technique is based on the down-conversion of the
incoming {Radio Frequency (RF)} signal (astronomical or from laboratory) resulting from its
mixing with the narrow signal of a local oscillator. The lower 
frequency allows to use lower noise amplifiers and, mostly, provides a
spectral resolution equal to the local oscillator line width, in
practice better than $10^{-8}$ of the RF frequency. Whereas
high sensitivity is needed for detecting the vanishingly weak
rotational line emission from rare molecular species, or from very
distant sources, high spectral resolution is crucial to identify with
certainty the carriers of detected astronomical lines. Similarly,
high spectral resolution is needed in the spectroscopic laboratory to
accurately measure the rotational line pattern, hence derive the
spectroscopic constants. 

{On the backend side, the availability of increasingly high-speed ADCs 
(Analog-to-Digital Converters) and powerful FPGAs (Field-programmable Gate Arrays)
have allowed the development of digital correlators that process increasingly wide
bandwidths, while keeping high spectral resolution.}   

\begin{figure}[h!]
\begin{center}
\includegraphics[width=10cm]{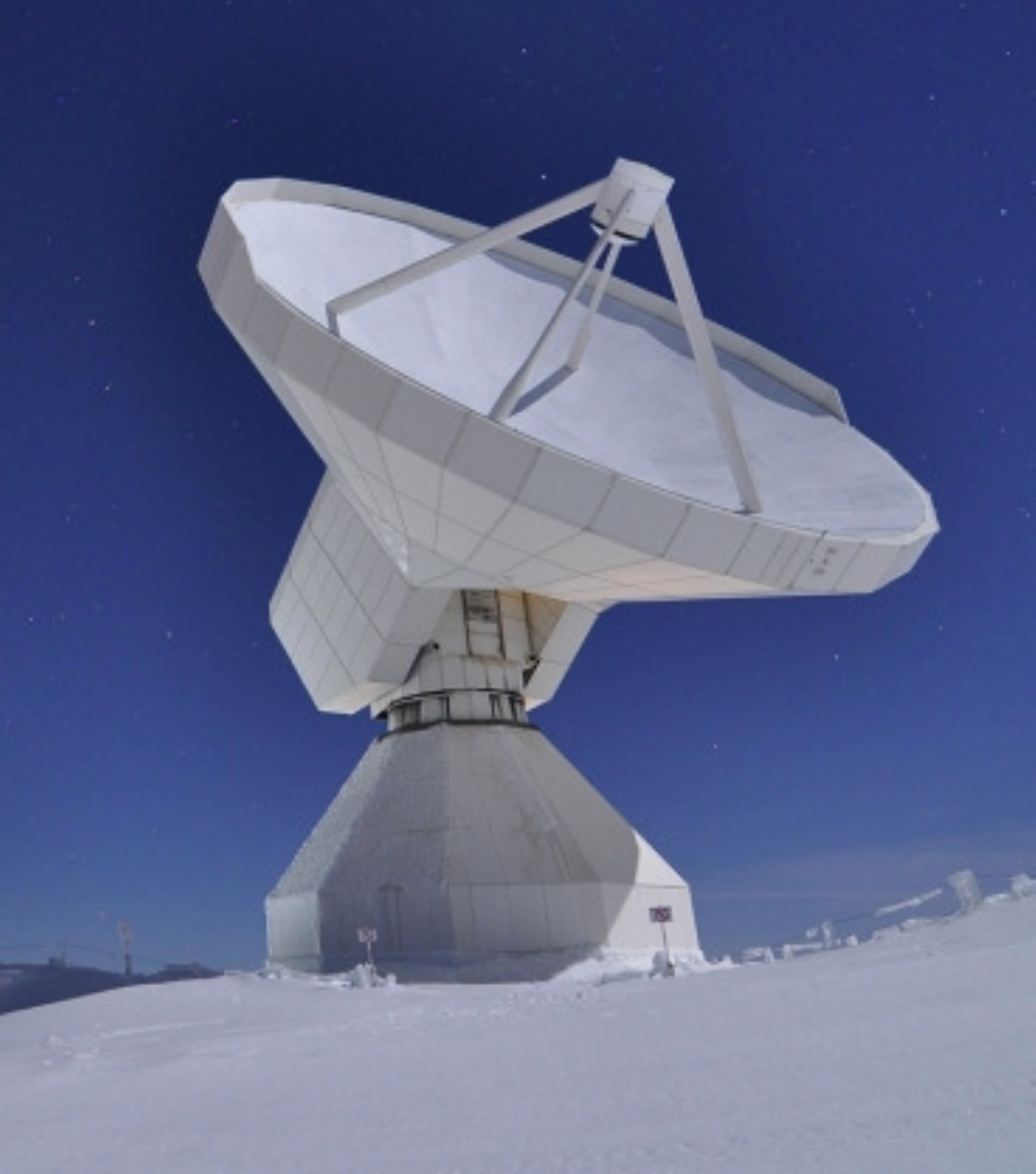}  
\end{center}
\caption{The IRAM 30-m diameter millimeter-wave telescope operating on Pico Veleta in the Sierra Nevada, Spain, at 2900 m altitude.
}\label{30m}
\end{figure}

\begin{figure}[h!]
\begin{center}
\includegraphics[height=17cm,angle=-90]{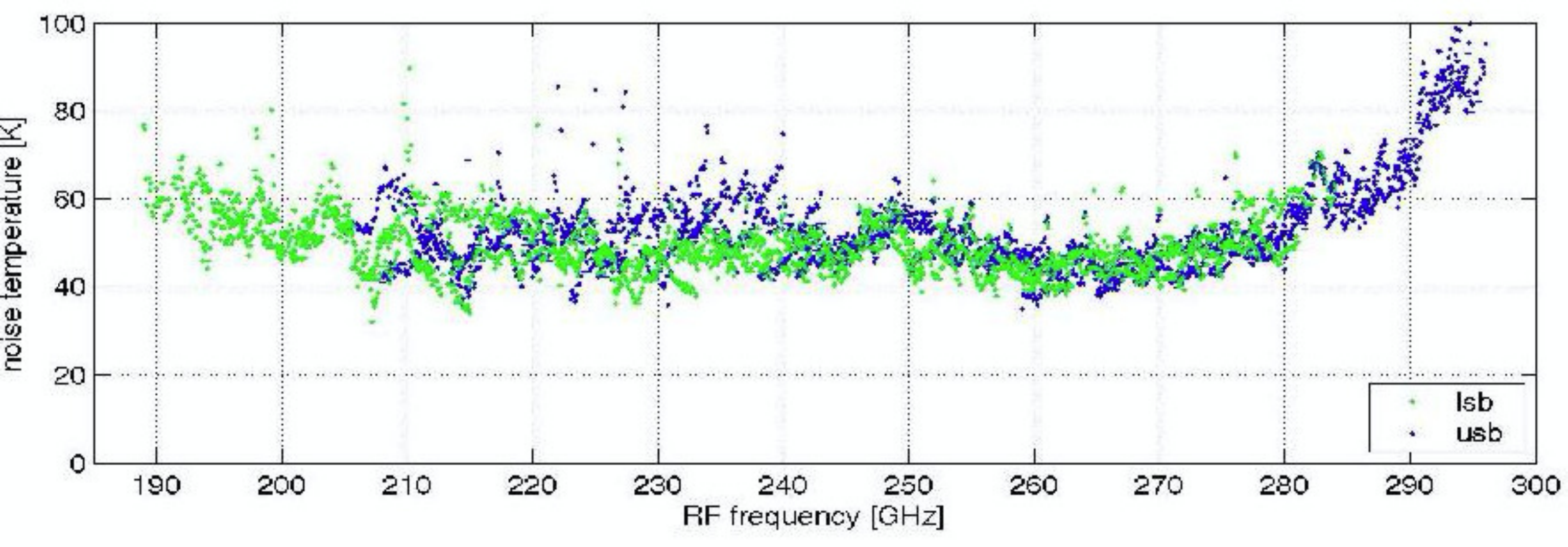}  
\end{center}
\caption{Noise temperature in the lower (green dots)  and upper (blue dots) sidebands of the 2SB 
SIS mixer receiver on the 30-m telescope -- credit: D. Maier, IRAM.
}\label{2SB-SIS}
\end{figure}

\begin{figure}[h!]
\begin{center}
\includegraphics[width=17cm,angle=0]{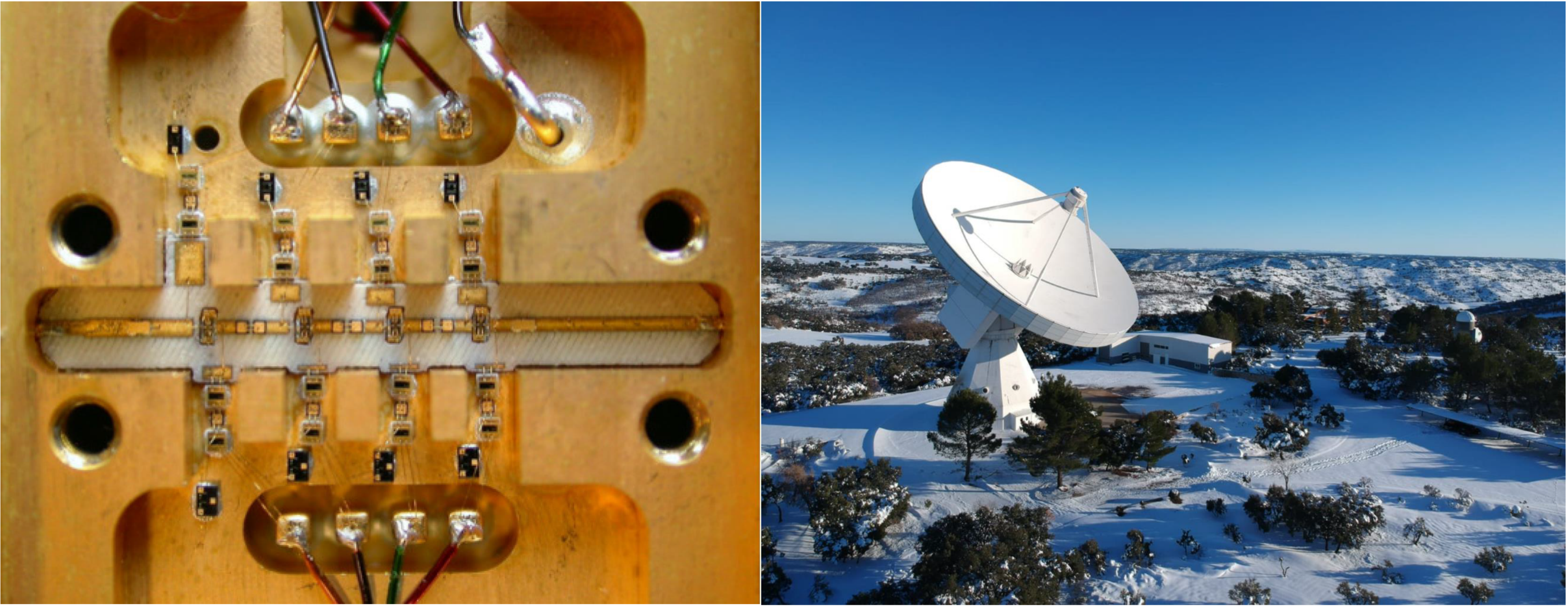} 
\end{center}
\caption{ ({\it Left}) The 4-stage Q-band cryogenic amplifier on the 40-m Yebes telescope (from Tercero et al.
2021). {\it (right)} The 40-m Yebes telescope --credit: Pablo deVicente}\label{Q-band-amp}
\end{figure}

The EMIR SIS junction receivers on the IRAM 30-m diameter telescope,
located at an altitude of 2900 m on Pico Veleta, near Granada, Spain
(Fig. \ref{30m}), allow to simultaneously observe a 32 GHz-wide band
with a spectral resolution of 200 kHz within the 73-375 GHz ($\lambda$
4 to 0.8 mm) atmospheric windows.  Their intrinsic noise within this
band varies between 2.5 and 5 times the quantum noise limit
(Fig. \ref{2SB-SIS}). The K-band HEMT receivers on the 100-m Green
Bank telescope, which can observe simultaneously a band of width 6 GHz
with a 66 kHz resolution, within the 26-40 GHz band, have a
r.m.s. noise of 7 times the quantum noise limit. The new Q-band HEMT
receiver on the Yebes 40-m telescope (Fig. \ref{Q-band-amp}; see
Tercero et al. 2021) can observe simultaneously 20 GHz, within the
31-50 GHz atmospheric window, with a 38 kHz resolution and, at
frequencies near 40 GHz, a r.m.s. noise around 4 times the quantum
limit.

Obviously, the telescope sensitivity also depends on the antenna size,
the accuracy of its reflecting surfaces (its aperture efficiency $\eta_A$ ) 
and the atmosphere transparency. Due to signal absorption by atmospheric water vapour, 
sensitivity is degraded at shorter radio wavelengths. This is why telescopes
operating below 2 mm (i.e. above 150 GHz) are built on high altitude sites. High frequency
telescopes must have more accurate surfaces to keep a high aperture
efficiency\footnote[1]{Aperture efficiency, $\eta_A$, according to Ruze's equation, 
scales with the r.m.s. surface error $\epsilon$ as 
$exp(-(4\pi\epsilon/\lambda)^2)$, which means that $\epsilon$ must be $\leq 50\mu$m to
achieve an efficiency $\eta_A\geq 0.5$ at a wavelength $\lambda\leq 1$mm.}. Large size,
high surface accuracy antennas located on high mountain sites are much more difficult
to build and need to be protected against harsh weather conditions. In practice,
this limits their diameter to about 30 m to operate efficiently at 1 mm wavelength {or below}.

In order to increase the effective area of a telescope intended to operate at wavelengths shorter than 2 mm, 
one chooses, rather than increasing the size of a single antenna dish, to build an 
array of several smaller dishes phased together, i.e to build an interferometer. 
Besides cost, a decisive advantage of a phased array of mobile antennas is that 
its angular resolution can be much better. The latter, which scales with the instrument 
largest dimension $D$ as $\lambda /D$, is a critical parameter in the radio domain. 

Another advantage of interferometers vs single-dish telescopes is that they
yield spectra with much flatter baselines, as they filter out intensity
fluctuations caused by amplifier instabilities and by the Earth atmosphere. This is
particularly valuable for the detection of the broad (broadened by Doppler effect) and very weak
lines, such as those observed in remote galaxies: As shown in Section 4.2, this makes 
it possible to detect, with proper integration time, molecular lines $10^5$ times
weaker than the receiver noise throughout a 8 GHz-wide spectrum without
removing any baseline. On a spectrum observed with a single dish, one would typically 
have to remove a high degree polynomial baseline, an operation that may affect
the detectability of broad lines.

\begin{figure}[h!]
\begin{center}
\includegraphics[width=17cm, angle=0]{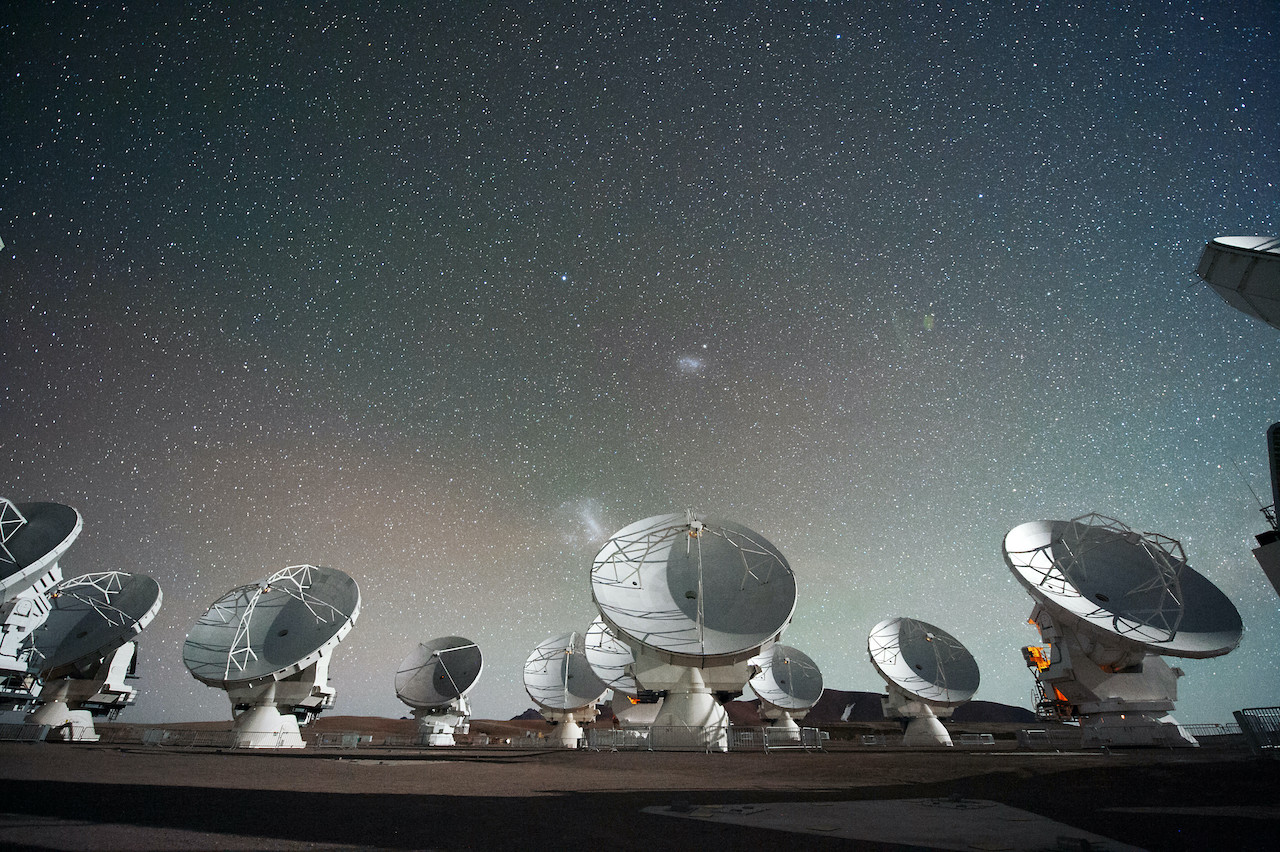}  
\includegraphics[width=17cm, angle=0]{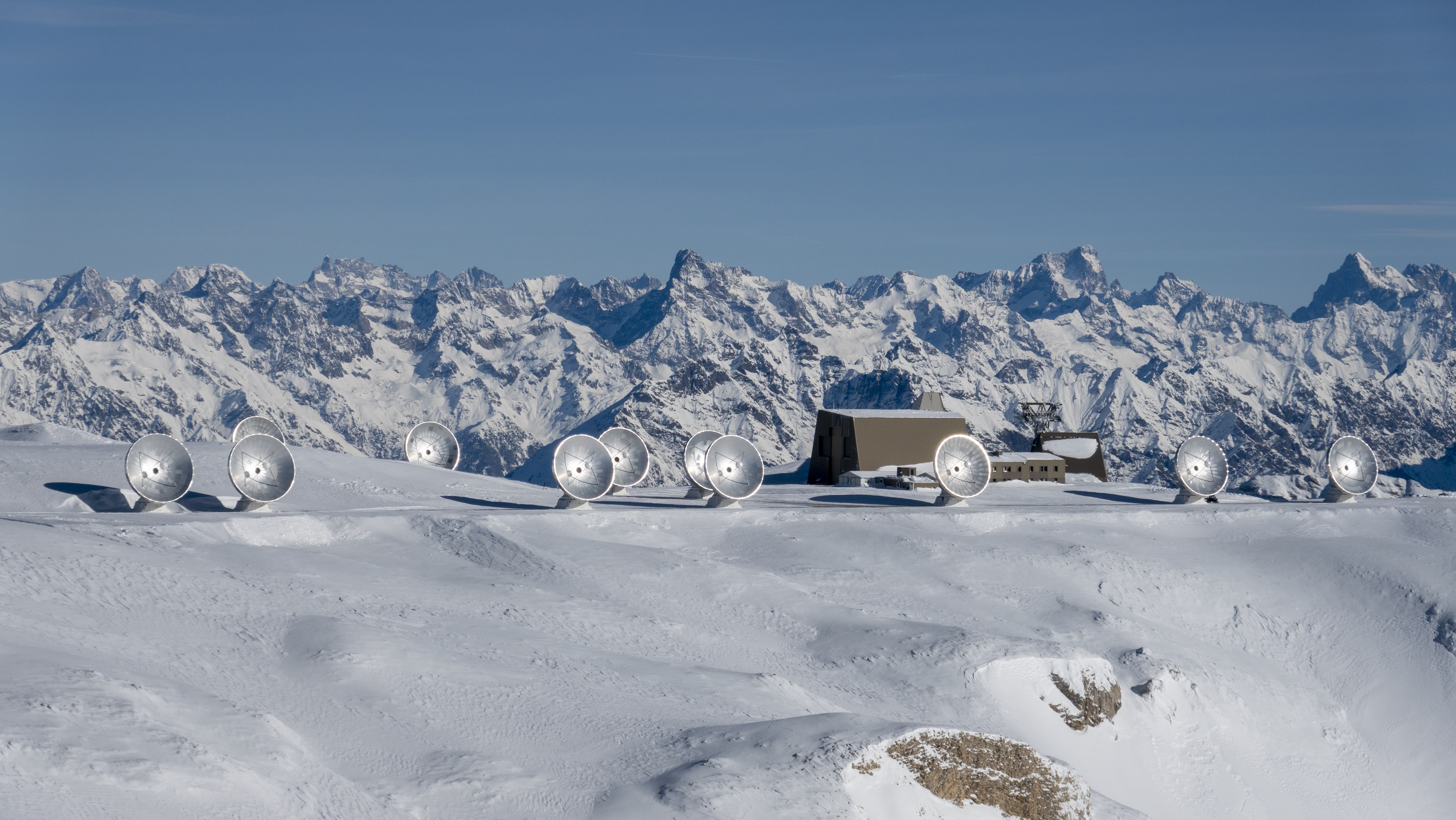}  
\caption{. 
{\it Upper)} Night view of the ALMA 12-m antenna array on the Chajnantor Plateau in the Chilean Andes at 5000 m altitude. 
The Magellanic Clouds appera as two white cloudlets in the night sky --credit: ESO.org. 
{\it Lower)} The NOEMA interferometer, operating on the Plateau de Bure in the Alps at 2550 m altitude. A 12$^{th}$
Antenna is being commissioned inside the hall and will join the array end of 2021 --credit: IRAM.   
}\label{NOEMA11}
\end{center}
\end{figure}

\begin{figure}[h!]
\begin{center}
\includegraphics[width=17cm,angle=0]{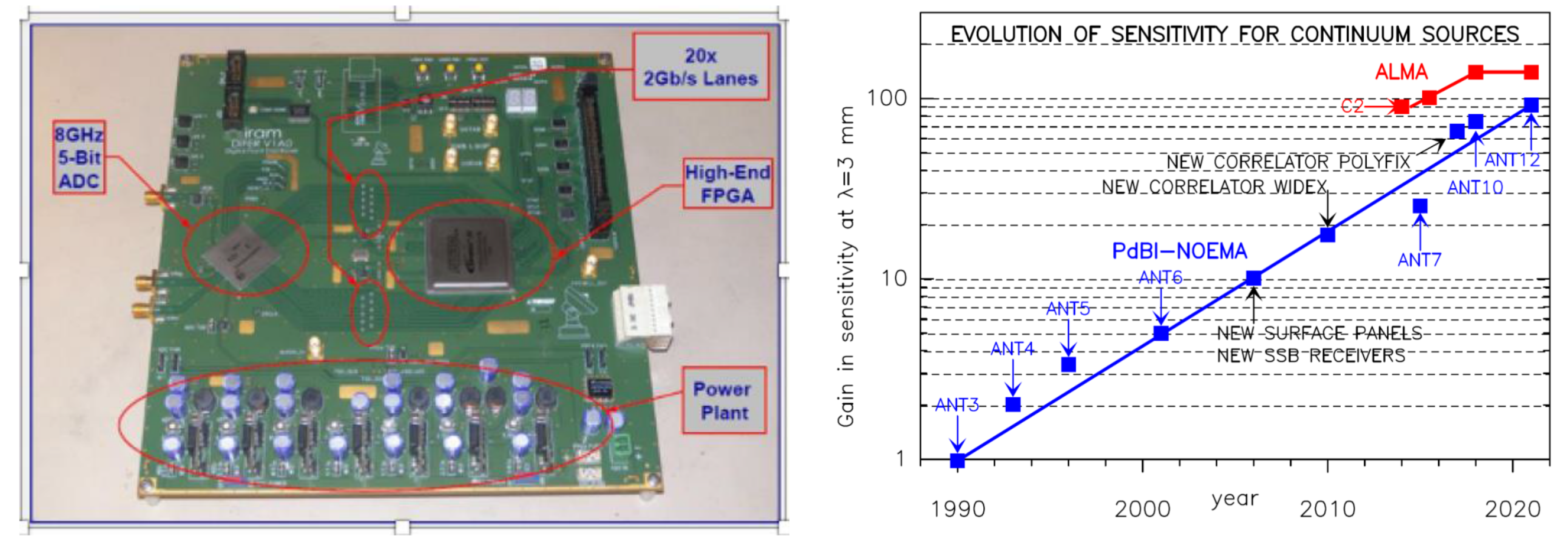}
\caption{{\it Left)} Fast sampling board of the IRAM Poyfix digital spectrometer. 
The board is based on a 16- Gsps, 5-bit ADC from the {\it e2v} company and a
{\it Stratix-IV} FPGA employed for later filtering and signal
processing. Eight such boards allow to simultaneously process the
signals from a 32 GHz-wide band detected by 12 antennas (see Gentaz et
al. 2012). Polyfix was implemented on NOEMA in 2017. In comparison,
the continuum correlator from the initial 3 antenna Plateau de Bure
array was only processing 10x50 MHz-wide sub-bands with a 4-bit
sampler.
{\it Right)}
Time evolution of the sensitivities of PdBI-NOEMA (blue squares)
and ALMA (red squares) for the detection at 3-mm wavelength of continuum
point sources (and for 3-mm of precipitable water vapour along the 
line-of-sight). Gains in sensitivity are relative to the year 1990 when the
first 3 antennas of the Plateau de Bure interferometer (PdBI)
were open to science observations. The ALMA
regular science observations started with Cycle 2 in 2014. Sensitivity
improvements for PdBI-NOEMA (two orders of magnitude in 30 years) 
resulted from a) an increase in effective collecting area
(addition of new antennas to the arrays and surface adjustments),
b) reduction of receiver noise and c) increase in bandwidths
(IF amplifiers and backend correlator).  End of 2021, the
r.m.s. noise figures of NOEMA (with 12x15-m diameter antennas) and
ALMA (43x12-m antennas) at 3-mm, after 8 h of integration, 
will be 7 and 3.5 $\mu$Jy respectively. Adapted from R. Neri.}
\label{Polyfix}
\end{center}
\end{figure}

The interferometers used for the line surveys discussed in the next
sections are ATCA, ALMA and NOEMA. ATCA is an array of 6 x 22-m
diameter antennas operating down to 7-mm wavelength (50 GHz) and
located at 240m altitude, $-30^\circ$ latitude, in New South Wales,
Australia.  The ALMA 12-m array consists of an array of 50 high
accuracy, 12-m diameter parabolic antennas, located at 5000 m altitude, $-23^\circ$
latitude, in the Atacama {desert} in Chile. It operates down to 0.3 mm
wavelength (950 GHz) thanks to the low water vapour content of the
atmosphere above the site. {This 12-m array is seconded by a smaller array of  
twelve 7-m antennas plus four 12-m antennas, ACA, also known as the Morito-san array 
(Fig. \ref{NOEMA11}).}
Finally, NOEMA {is upgraded end of 2021 to a 12 x 15-m diameter antenna array;
it is} located on a 2550 m altitude plateau in the southern Alps in France and
operates down to 1 mm wavelength with 12 antennas (and currently 0.8mm, or 375 GHz, 
with 6 antennas). 
NOEMA is the largest
mm-wave array in the Northern Hemisphere (Fig. \ref{NOEMA11}). Its SIS junction receivers
and its wide band XF Fast Fourier Transform correlator make it
particularly well suited for extragalactic spectral surveys, since it
can instantaneously cover a 32 GHz-wide bandwidth with 2 MHz-wide (2.6
kms$^{-1}$ at 230 GHz) spectral channels {(see Fig. \ref{Polyfix})}.

\vspace{1.0cm}
\section{Complex organic molecules in the dark cloud TMC-1}
Cold dark clouds have turned up to be {amazingly} rich
chemical laboratories, able to synthesize in situ a great
variety of molecules. Most of the identified species are neutral molecules,
although several cations, essentially protonated forms
of abundant closed-shell species (Agundez et al. 2015a,
Marcelino et al. 2020, Cernicharo et al. 2020a, 2021a,b), but also a
few hydrocarbon and nitrile anions, have also been observed
(Cernicharo et al. 2020b). Molecular
ions have low abundances, typically below 10$^{-10}$ relative to
H$_2$, due to their high reactivity.

TMC-1 is a cold pre-stellar core located within Heiles' Cloud 2 in the
Taurus constellation (Cernicharo \& Gu\'elin 1987).  Target of very
many studies (see Fuente et al. 2019, and references therein) it 
{was early on} recognized as a rich molecular source. Within the Taurus
cloud complex, several dark cores present similar physical
characteristics (Cernicharo et al. 1984), 
but TMC-1 stands out as
the one where the largest amount of carbon-bearing molecular species
has been detected.  

TMC-1 presents an interesting carbon-rich
chemistry that leads to the formation of long carbon-chains, radicals,
ions and neutral closed-shell molecules such as cyanopolyynes (see
Cernicharo et al. 2020b,c and references therein).  This cold dark
core also hosts a number of nearly saturated species, such as
CH$_3$CHCH$_2$, more typical of hot star-forming {cores} (Marcelino et
al. 2007). A first polar benzenic ring, benzonitrile C$_6$H$_5$CN, has been
detected in this object (McGuire et al. 2018a), although benzene itself
is only observed in post-AGB star envelopes. In spite of this
carbon-dominated chemistry, a number of carbon chains containing
oxygen have been detected in TMC-1: CCO (Ohishi et al. 1991), C$_3$O
(Matthews et al. 1984), HC$_3$O$^+$ (Cernicharo et al. 2020a),
{HC$_3$O and C$_5$O (Cernicharo et al. 2021k)}
HC$_5$O, and HC$_7$O (Cordiner et al. 2017, McGuire et al. 2017; {see
also Cernicharo et al. 2021k)}. The
path leading to the formation of HC$_5$O and HC$_7$O remains a
mystery {(Cernicharo et al. 2021k)}, all the more that the related chains {C$_4$O, HC$_4$O, and
HC$_6$O} have not been found (HC$_2$O is nethertheless observed in
other cold dense clouds -- Agundez et al.  2015b). Other O-bearing
species, more typical of hot cores and {corinos}, such as {
CH$_3$OH, C$_2$H$_3$CHO}, C$_2$H$_3$OH, HCOOCH$_3$ and CH$_3$OCH$_3$,
have also been identified in TMC-1 mm-wave spectra (Agundez et
al. 2021a).

Since L\'eger and Puget's assertion (1984) 
that the carriers of the
unidentified infrared bands must be polycyclic aromatic hydrocarbons
(PAHs), efforts have been devoted to
understand how those species may form in space.  
Cherchneff et al. (1992) proposed that the
envelopes surrounding carbon-rich AGB stars could be PAH
factories. Along this line, the detection of benzene in the C-rich
protoplanetary nebula CRL\,618 (Cernicharo et al. 2001) suggested a
bottom-up approach in which the small hydrocarbons formed during the
AGB phase, such as C$_2$H$_2$ and C$_2$H$_4$, interact with the
ultraviolet (UV) radiation emitted by the star when it evolves towards
a white dwarf (Woods et al. 2002, Cernicharo 2004).  Alternately, in a
top-down approach, PAHs would result from the stripping of graphite
grain surfaces chemically processed or irradiated by stellar UV
(Pilleri et al. 2015, Mart\'inez et al. 2020).  Hence, it is startling
to discover that a cold dark cloud, shielded from UV and devoid of
internal energy sources, hosts not only very long C-chain molecules
such as HC$_9$N, but also C-cycles and aromatic rings.

\begin{figure}[h!]
\begin{center}
\includegraphics[width=15cm]{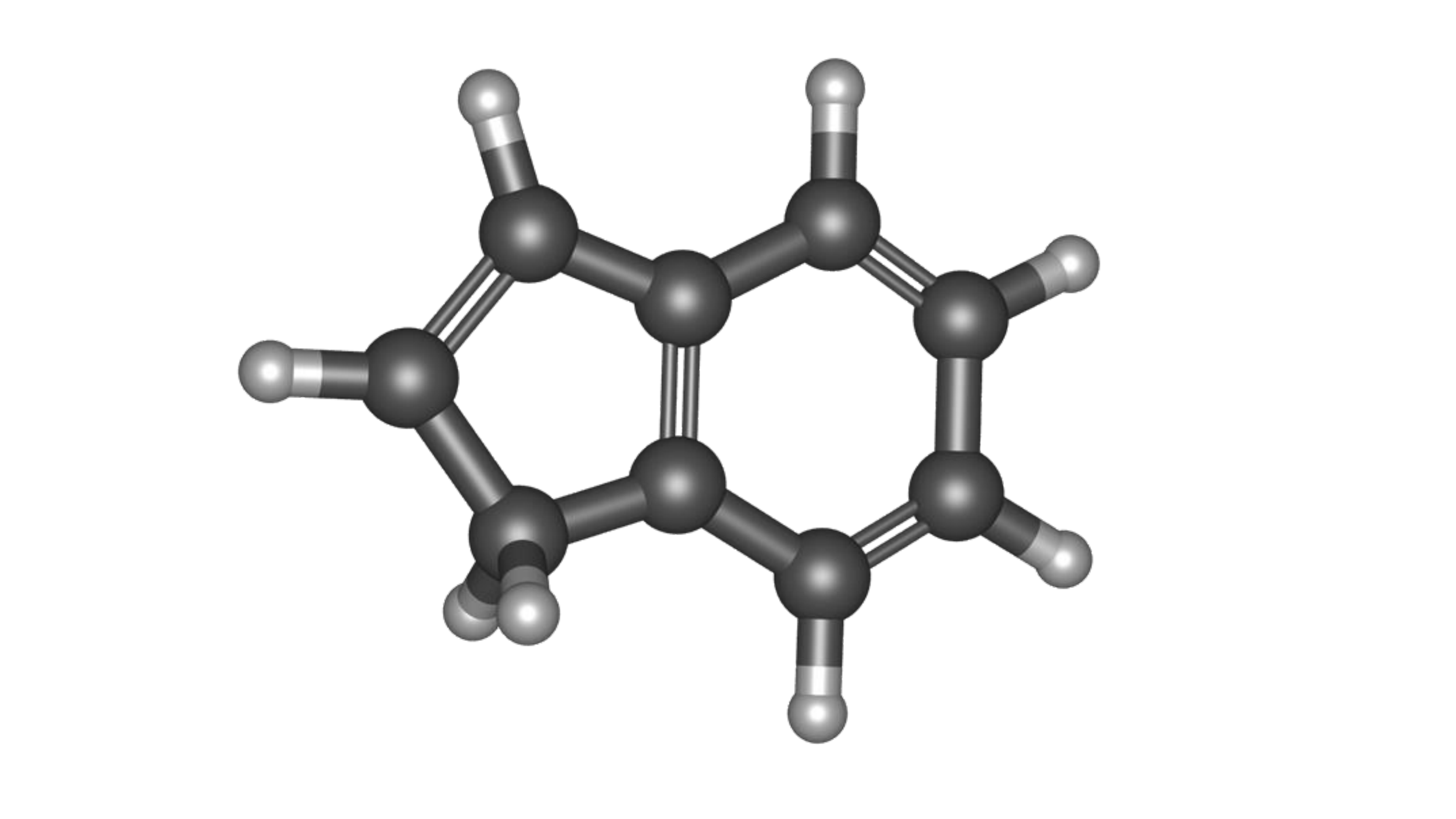} 
\caption{Indene structure (from Cernicharo et al. 2021d).} 
\label{indene_structure}
\end{center}
\end{figure}

\begin{figure}[h!]
\begin{center}
\includegraphics[width=17.0cm]{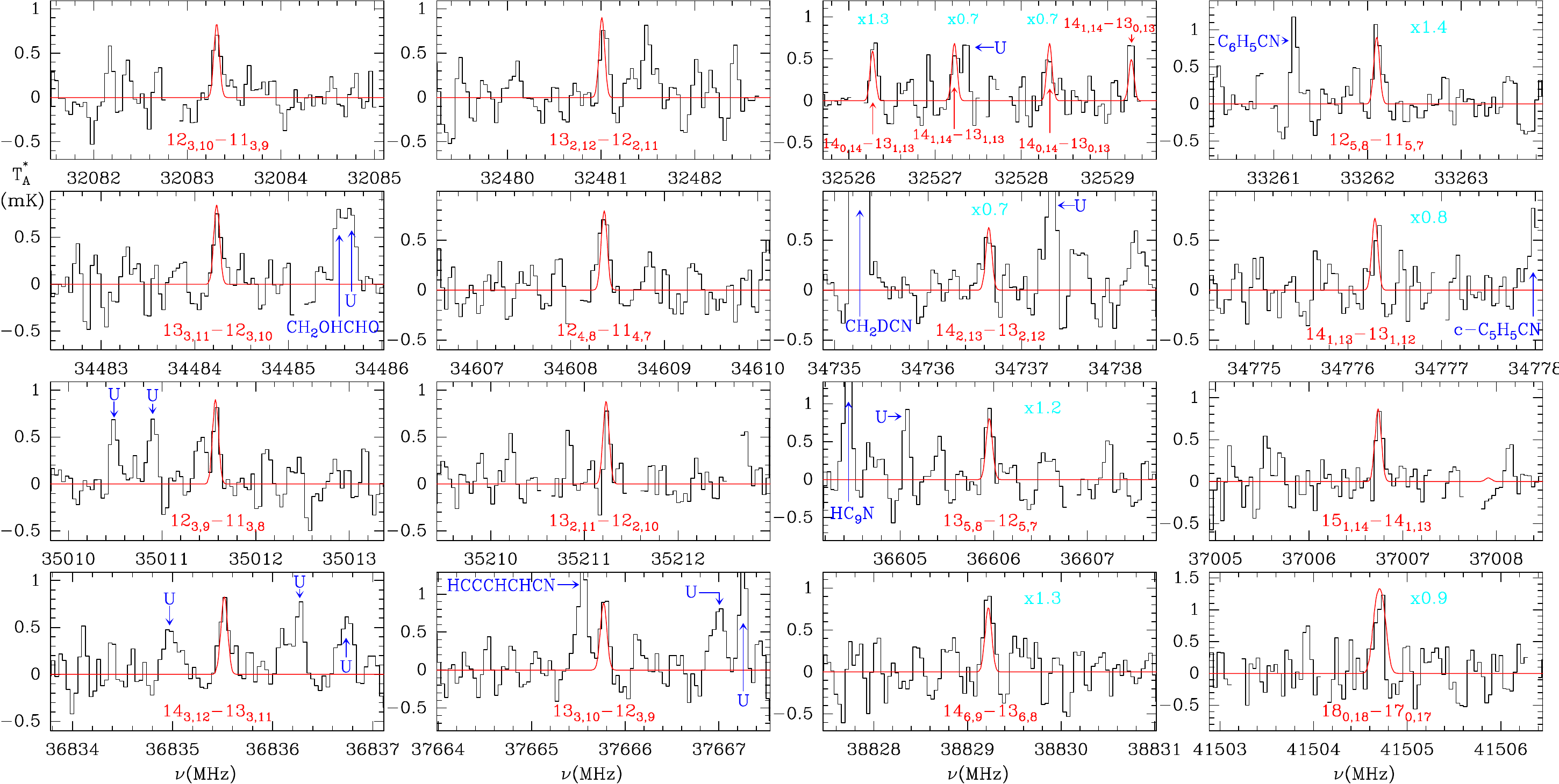}  
\caption{Indene transitions observed {towards TMC-1} with the Yebes 40-m radio telescope by Cernicharo et al. 2021d.}
\label{indene_spectra}
\end{center}
\end{figure}

\begin{figure}[h!]
\begin{center}
\includegraphics[width=17.0cm]{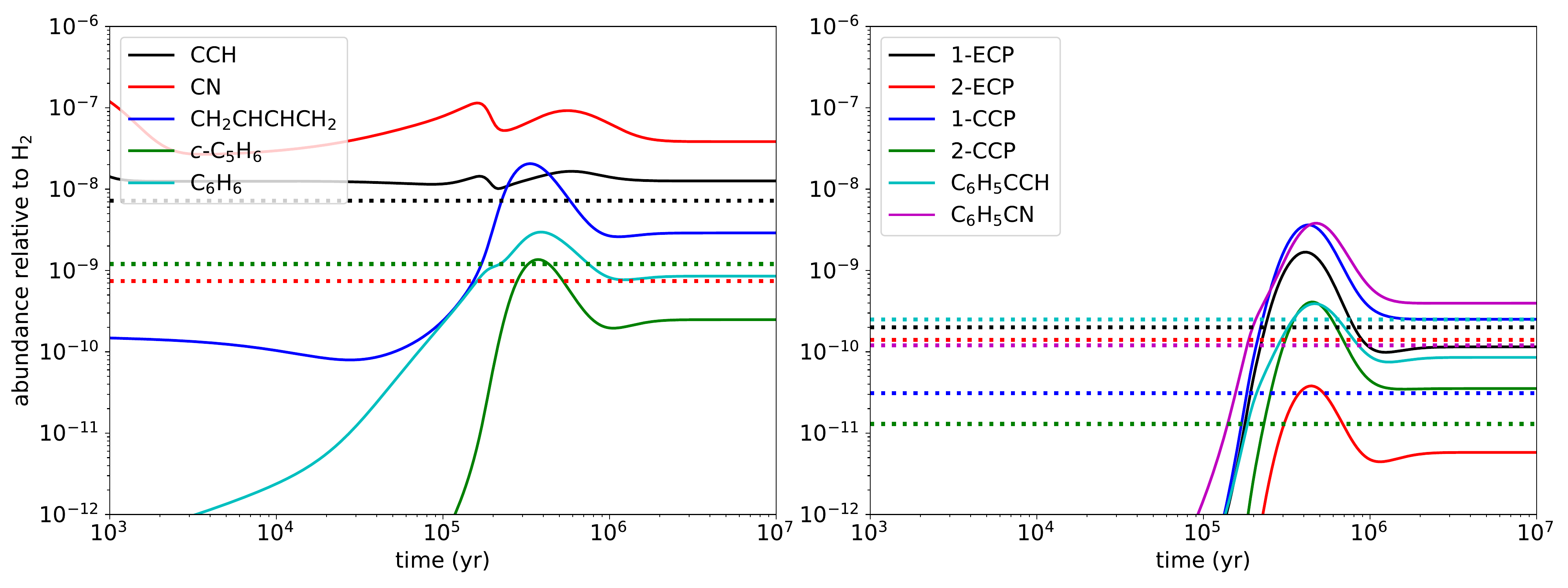}  
\caption{Calculated abundances of CCH and CN derivatives of $c$-C$_5$H$_6$ 
and C$_6$H$_6$ (right panel) and of their precursors (left panel). The 
horizontal dotted lines correspond to the abundances observed in TMC-1. {The species
1/2-ECP correspond to two isomers of ethynyl cyclopentadiene, while the species
1/2-CCP correspond to
two isomers of cyano cyclopentadiene (from Cernicharo et al. 2020g)}.}         
\label{fig_abun}
\end{center}
\end{figure}

This exciting result mainly stems from two on-going line surveys of
TMC-1 between 20 and 50 GHz: GOTHAM (\textbf{G}BT
\textbf{O}bservations of \textbf{T}MC-1: Hunting \textbf{A}romatic
\textbf{M}olecules; McGuire et al. 2018) and QUIJOTE ({\textbf{Q}-band
\textbf{U}ltrasensitive \textbf{I}nspection \textbf{J}ourney in the
obscure \textbf{T}MC-1 \textbf{E}nvironment; Cernicharo et
al. 2021c). {Both surveys follow the pioneering work of Kaifu et
al. (2004), who first surveyed the whole 8.8 to 50 GHz frequency
spectrum in that source with the Nobeyama 45-m telescope. N. Kaifu,
S. Saito and their colleagues identified in the surveyed band 414
lines pertaining to 38 molecular species. The Nobeyama survey data
were subsequently used by McGuire and collegues (2018a) to find,
through spectral stacking, a hint of the presence of benzonitrile, a
presence superbly confirmed by the GOTHAM survey.}

The GOTHAM and QUIJOTE surveys represent a major leap in sensitivity with respect to previous works. They use the 100-m 
diameter GBT and the recently completed 40-m Yebes radio telescope, that benefit 
from higher effective areas and, mostly, from lower noise and broader bandwidth receivers.   
The teams of both surveys reported the detection of several cyclic molecules. 
The GOTHAM team mostly used a spectral stacking technique (consisting,
for a given polar molecule, in a weighted average of the very many
rotational transitions covered by the survey) to achieve their
detections. The QUIJOTE team, owing to the higher (sub milli-kelvin)
sensitivity of its survey, was able to detect individual lines without
the need to rely on blind line stacking (see
Fig. \ref{indene_spectra}). Among the QUIJOTE detections, we stress
those of indene, the first double-cycle detected in space 
(Fig. \ref{indene_structure} and \ref{indene_spectra} and Cernicharo et
al. 2021c; see also Burkhardt et al. 2021), of cyclopentadiene
(Cernicharo et al. 2021d), and $ortho$-benzyne (Cernicharo et
al. 2021c). 
Detections reported by the {GOTHAM and QUIJOTE teams} include ethynyl and
cyano derivatives of {cyclopentadiene, benzene, and naphthalene (McCarthy et
al. 2021, Lee et al. 2021, Cernicharo et al. 2021j; see also Cernicharo et al. 2021f)}.

It is unlikely that such complex molecules and cycles arose from a
reservoir pre-existing to the formation of the TMC-1 core: they could
have hardly survived in the diffuse gas transparent to UV radiation.
McGuire et al. (2021) propose a reasonable chemical network starting
with the phenyl radical that may explain the observed abundances of
cyanonaphthalene, but the chemical routes leading to this radical and benzene itself
remain unclear. An {\it in situ} formation mechanism for benzene must
involve abundant hydrocarbons containing from 2 to 4 carbon
atoms. Moreover, those hydrocarbons should react and form C-rings in a
mere 2 or 3 steps to be able to form enough benzene or phenyl
radicals.

The propargyl radical, CH$_2$CCH, has recently been found in TMC-1 by
Ag\'undez et al. (2021b) with an abundance close to 10$^{-8}$ relative
to H$_2$. In addition, complex hydrocarbons such as vinyl and allenyl
acetylene have also been observed with fairly large abundances
(Cernicharo et al. 2021e,f).  These hydrocarbons may hold the key to form the
first aromatic rings, from which larger PAHs can grow. In a recent
article reporting the detections of ortho-benzyne and the ethynyl
derivatives of cyclopentadiene (Cernicharo et al. 2021c,g), the
authors reproduce reasonably well the observed abundances of C-cycles
with reactions between the radicals CH, CCH, C$_4$H, H$_2$CCCH,
CH$_2$,... and the C-chains CH$_3$CCH, H$_2$CCCH, CH$_2$CHCCH, and
H$_2$CCCHCCH (see Fig. \ref{fig_abun}).  This requires, however, to
reduce the abundance of oxygen to a level close to that of carbon: the
presence of free oxygen atoms, which efficiently react with the
hydrocarbons, would strongly reduce the yield of five or six C-atoms
rings, hence impede the formation of PAHs. A similar result was
reported for other species by Fuente et al. (2019).

In addition to hydrocarbons (closed shells and radicals) and their
cyano and ethynyl derivatives, TMC-1 revealed itself as an efficient
sulfur factory, through the identification of a number of complex
sulfur-bearing species in the QUIJOTE survey: HC$_3$S$^+$, HCCS, NCS,
H$_2$CCS, H$_2$CCCS, C$_4$S, HCSCN, and HCSCCH (Cernicharo et
al. 2021h,i).

Most remarkably, in just a couple of years, The GOTHAM and QUIJOTE surveys
have increased the number of known interstellar molecules from 204
to 256 and revealed the great potential of TMC-1 for further investigations.

\vspace{1.0cm}
\section{Molecules in the distant Universe}

How much can we learn about the interstellar gas in nearby galaxies
and farther out, in the remote parts of the Universe? Observing
increasingly distant sources means travelling back in time. 
Because the Universe is expanding, the radiation emitted 
by distant sources (and the spectral lines it may contain,
e.g. Ly$\alpha$, CII and MgII lines) appears to us redshifted. 
Models of the Universe expansion tell us the relation between
redshift and distance. Except for the 2.7 K Cosmic Background, the
remotest sources we can observe are the very first quasars, the bright 
radio/infrared quasi-stellar sources that appeared at the center of 
the primordial galaxies during the epoch of reionization, some 700 million years 
after the Big Bang. The most distant quasar identified as of 
today, {J0313-1806, has a redshift $z\simeq 7.64$ when the universe was only 680 million 
light-years old (Wang et al. 2021).}

Whereas the lines observed in {optical} spectra teach us about these
remote quasi-stellar objects' redshift, age and elemental composition, they are far
too broad to inform us on the dynamics of the emitting gas  {(see e.g. Wildy \& Czerny 2017)} 
and, mostly, on its molecular and isotopic compositions. 
Those latter can only be studied at
{sub-millimeter and} millimeter wavelengths,  
preferably in circumstances that compensate for the {extreme} line weakness. 

The most favourable case is when the line-of-sight to the remote
quasar is intercepted at halfway by a massive galaxy or a cluster of
galaxies (halfway is the optimum distance for a magnification of the
quasar radiation by gravitational lensing).  If the quasar is
radio-loud, the gas of the intervening galaxy may absorb the quasar
continuum radiation, causing the rotational line pattern of the polar
molecules to appear in absorption against the magnified background continuum. The
gas disks of galaxies are thin, so that the absorption lines are fairly
narrow and their carriers can be unambiously identified.  On the same
mm spectrum, but at longer wavelengths (i.e. at higher redshift), one
may detect in emission, magnified by the lens, molecular lines arising
in the quasar itself.

{The Planck/HIFI, Hershell/HerMES, H-ATLAS and South Pole Telescope 
far-infrared/sub-millimetre surveys show that a fair fraction (several percent)    
of the high-z sub-millimetre-bright galaxies are magnified by 
gravitational lensing (see Ade et al. 2016, Spilker et al. 2016, as well as references 
in Hodge and da Cunha, 2021). 
Nonetheless, and except for the relatively strong CO and H$_2$O lines (Pensabene et al. 2021, Yang et al. 2016), 
rare are the cases where the magnification is large enough to allow a clear detection 
of molecular lines from the intervening galaxies, or from the background quasar itself.
 
Besides the detections in a handful of lensed quasars (e.g. APM08279+5255) of a few lines of HCN, HNC and HCO$^+$  
(Garcia-Burillo et al. 2006, Gu\'elin et al. 2007), one has essentially to rely on source stacking methods 
to barely detect more molecular species (Spilker et al.2014, Reuter et al. 2020).}
A couple of sources, however, are lucky exceptions: a field galaxy on the line-of-sight
to the bright radio-loud quasar PKS1830-211, and a strongly lensed quasi-stellar object, H1413+117, dubbed 
{\it the Cloverleaf}. We now focus on those two remarkable objects.}
\vspace{0.3cm}
    
\subsection{Molecules inside an arm of a spiral galaxy at redshift
z=0.89}

\begin{figure}[h!]
\begin{center}
\includegraphics[width=13cm,angle=0]{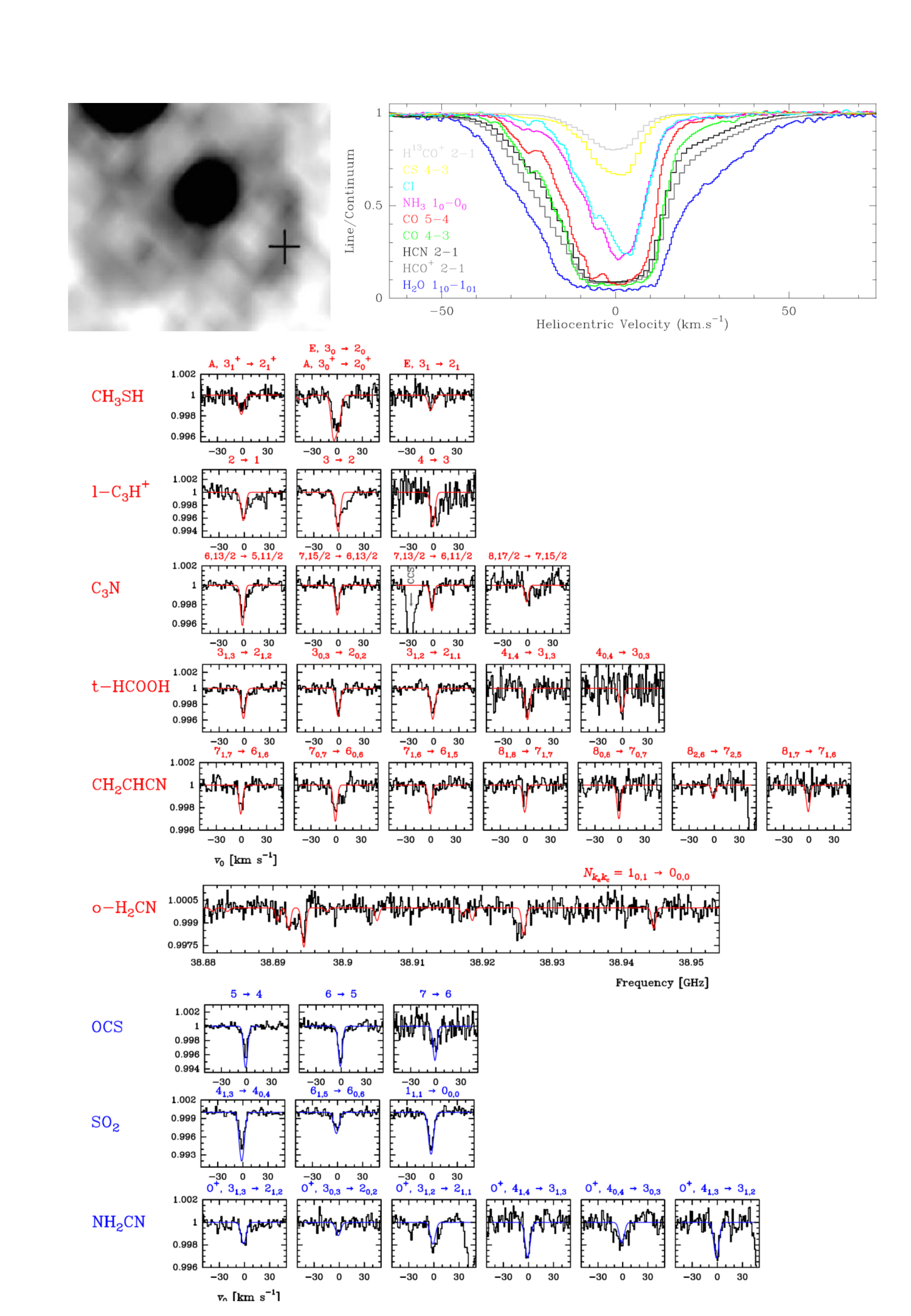}  
\caption{{\it Upper left:} { The spiral galaxy at z=0.89 intercepting the line
of sight to the quasar PKS 1830-211, as seen by the Hubble satellite in
the I band (Winn et al. 2002). The cross shows the position where the molecular lines are
observed in absorption in the SW spiral arm.  {\it Upper right:}
Profiles of the strongest absorption lines observed by ALMA in the SW
arm (Muller et al. 2014). Abscissa is the gas velocity in the source
frame (i.e. after correction for the redshift of the source
z=0.885875). Ordinate is the line to continuum ratio, where the
continuum arises from the SW image of the background quasar.
{\it Bottom panels:} Absorption lines observed toward the SW arm with the Yebes 40-m 
telescope. The species identified for the first time in an extragalactic source are highlighted 
in red. The molecules highlighted in blue are those that were identified for the first time
in this source, but observed previously in other extragalactic sources. The y-axis is the normalised 
intensity to the total (NE+SW) continuum level (from Tercero et al. 2020).
}
}\label{spiral}
\end{center}
\end{figure}

With a continuum flux of 2-3 janskys at 3-mm, the quasar
PKS~1830-211 (redshift z=2.5, Lidman et al. 1999) is the
brightest lensed source at millimetre wavelengths. Its line-of-sight
is intercepted by two galaxies: the nearest to us is responsible for
HI absorption at redshift z=0.19; the other, which appears on HST
images as a face-on spiral galaxy, gives rise to HI and molecular line
absorption at z=0.89 (see e.g. Muller \& Gu\'elin 2006, Combes et al. 2021). The galaxy acts
as gravitational lens that splits the quasar image into two compact
radio images, respectively located 0.6$''$ NE and 0.4$''$ ($\simeq 6$
kilo light-years) SW of the galaxy center.
The compact SW radio image lies just behind a spiral arm visible on
the I-band HST image (Fig. \ref{spiral}). 

The millimeter wave spectrum observed along both NE and SW lines-of-sight displays a dense pattern 
of absorption
lines. The SW one is particularly rich and not unlike those observed in the arms intercepting the
line-of-sight to the HII region SgrB2, close to the Milky-Way center,
or those inside the central starburst of the nearby Sc spiral galaxy NGC 253 (Martin et al. 2006). A
detailed analysis of the line intensities and profiles, as well as of
the chemistry of their carrier molecules, allows us to estimate the
physical conditions and morphology of the clouds intercepting both
lines-of-sight (Muller \& Gu\'elin 2008, Muller et al. 2014). The SW
line-of-sight crosses a number of mostly cold ($\leq 40$ K)
relatively dense clouds ($n_{\rm H_2} \geq 10^4$ H$_2$ molecules per
cm$^3$), surrounded by a diffuse
envelope. The NE one encounters only diffuse clouds ($n_{\rm H_2} \leq
10^3$ cm$^{-3}$) transparent to stellar UV radiation.
   
A total of 54 molecular species (not counting isotopologues) have been
identified in the parts of the spectrum surveyed to date: a 20 GHz-wide band
at 7-mm wavelength with the ATCA interferometer and the Yebes 40-m
telescope, a 8 GHz-wide band at 3-mm with NOEMA, and a 38 GHz-wide band between 3-mm and
0.6-mm with ALMA (Muller et al. 2011, 2014, 2016, 2017 and references
therein, Tercero et al. 2020).

These species are light hydrides (HF, CH, CH$^+$, CH$_3^+$, ND,
NH$_2$, NH$_3$...), small hydrocarbons (CCH, C$_3$H, C$_4$H, {C$_3$H$^+$, H$_2$CN, C$_3$N,} ...),
sulfur-bearing species (SH$^+$, SO, SO$_2$, NS...) but also heavier,
more complex organic molecules with up to 7 atoms {(see Fig. \ref{spiral})}: aldehydes
(formaldehyde and thioformaldehyde, acetaldehyde CH$_3$CHO), imines
(CH$_2$NH methanimine), amines (CH$_3$NH$_2$ methylamine), 
{cyanides (CH$_2$CHCN, vinyl cyanide)}, amides
(formamide NH$_2$CHO, urea (NH$_2$)$_2$CO), one alcohol (methanol
CH$_3$OH), {and its thioalcohol (CH$_3$SH, methyl mercaptan)}, and one organic acid 
(formic acid HCOOH).  Their detection 
suggests that all the ingredients necessary for a rich prebiotic
chemistry are present in the gas, although the physical conditions
(density, temperature) necessary for starting such a chemistry are
ways from being fulfilled even in the dense and hot prestellar cloud
cores. {Nevertheless, the chemical composition of the gas in this distant
object is not very different from that of interstellar clouds in the
Milky-Way.} 
 
The basic ingredients of Miller-Urey experiment (H$_2$, H$_2$O, CH$_4$ and NH$_3$)
and of closely related experiments (CO, CO$_2$, H$_2$S, SO$_2$) are all present
there (the nonpolar species H$_2$, CH$_4$ and CO$_2$ cannot be observed in the radio
domain). Also present are some emblematic intermediate products of
these experiments: hydrogen cyanide and isocyanide, formaldehyde,
urea, acetylene, cyanoacetylene, formic acid and, mostly, formamide
the key to the peptide bond. As shown by the experimental work and
ab-initio calculations reported by Ferus et al. (2018), formamide is a possible source
for the synthesis of amino acids and of the four RNA nucleobases (uracil,
cytosine, adenine and guanine).

\subsection{Molecules in a young quasar at redshift z=2.6}  

The Cloverleaf Quasar (H1413+117  is the archetype of gravitationally lensed 
quasars and one of the most luminous and best studied objects in the distant, 
hence young Universe. The quasar redshift (z=2.55786) implies a distance of 11 
billion light-years from the Sun {and an age of about 2  Gyr after the appearance of
the very first stars, 700 million years after the Big Bang. 
Located at the center of a starburst galaxy, its 
radiation, viewed from the Earth, is magnified by a factor as large as 11 (according to  
models of Kneib et al. 1998, Venturini \& Solomon 2003) by 2 galaxies, with redshifts 
z$\simeq 1$ and 2, respectively, that intercept the line-of-sight and split the quasar image into 
4 distinct components Fig. \ref{Chandra}}).  

The Cloverleaf quasar and its host galaxy have been studied in the X-ray,  optical and radio domains,
and more particularly at sub-millimetre wavelengths in the continuum and the lines of CO and HCN. 
The far infrared/sub-millimetre emission appears to arise from an extended (3 light-year radius) disk of gas and dust 
rotating around the active galactic nucleus (or AGN -- Barvainis et al. 1997, Alloin et al. 1997, 
Solomon et al. 2003, Bradford et al. 2009).

\begin{figure}[h!]
\begin{center}
\includegraphics[width=17cm]{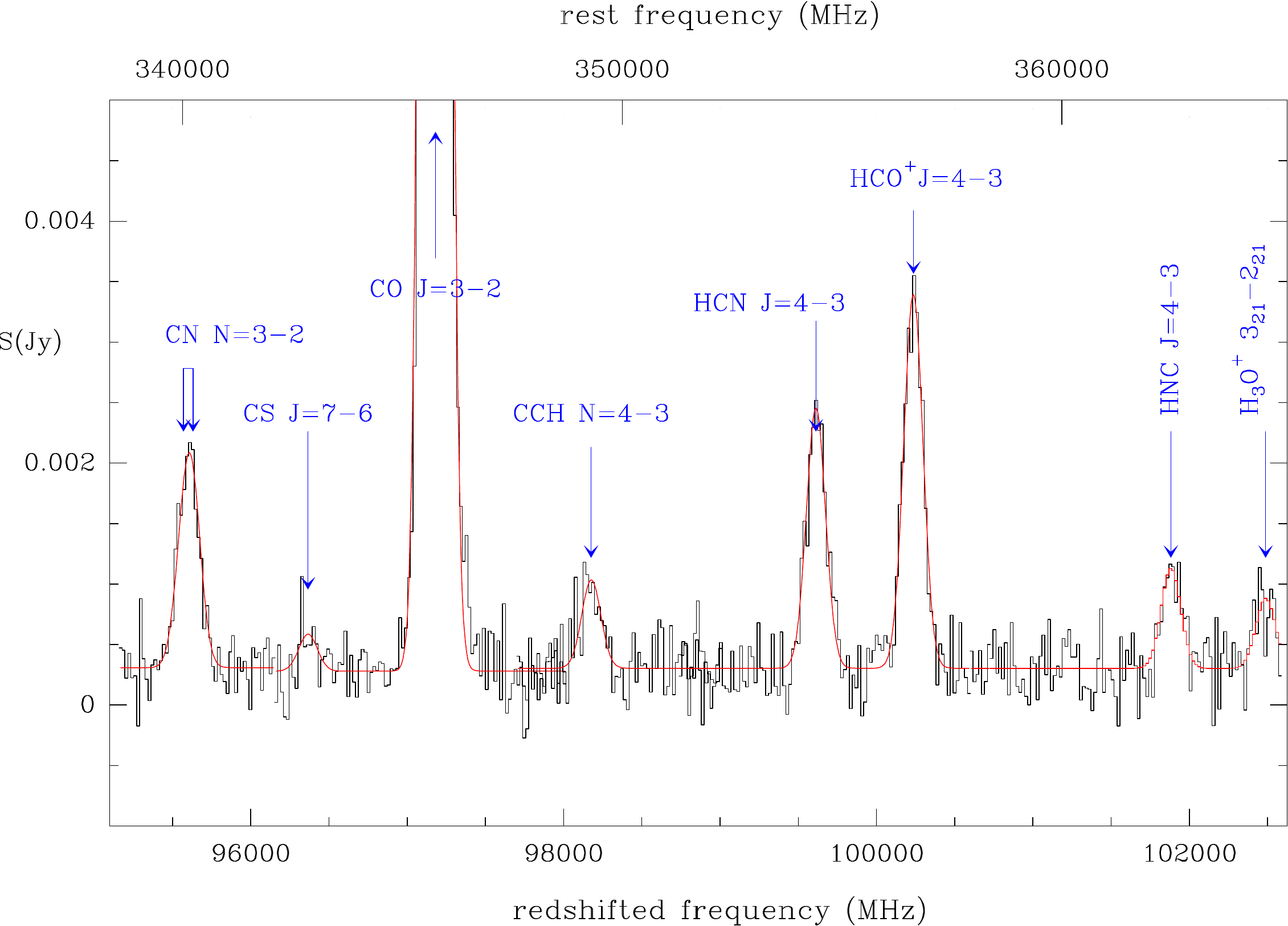}
\caption{ALMA 3-mm spectrum of the Cloverleaf Quasar covering the
N/J=3-2 lines of CN and CO, and the J=4-3 lines of HCN and
HCO$^+$. Ordinate scale is line flux in janskys. Abcissa is
observed frequency in MHz. Upper scale is rest frequency in the quasar frame. 
On-source integration time is 14h. -- Gu\'elin et al. {\it
in prep.}}\label{Clover-3mm}
\end{center}
\end{figure}

\begin{figure}[h!]
\begin{center}
\includegraphics[width=17cm]{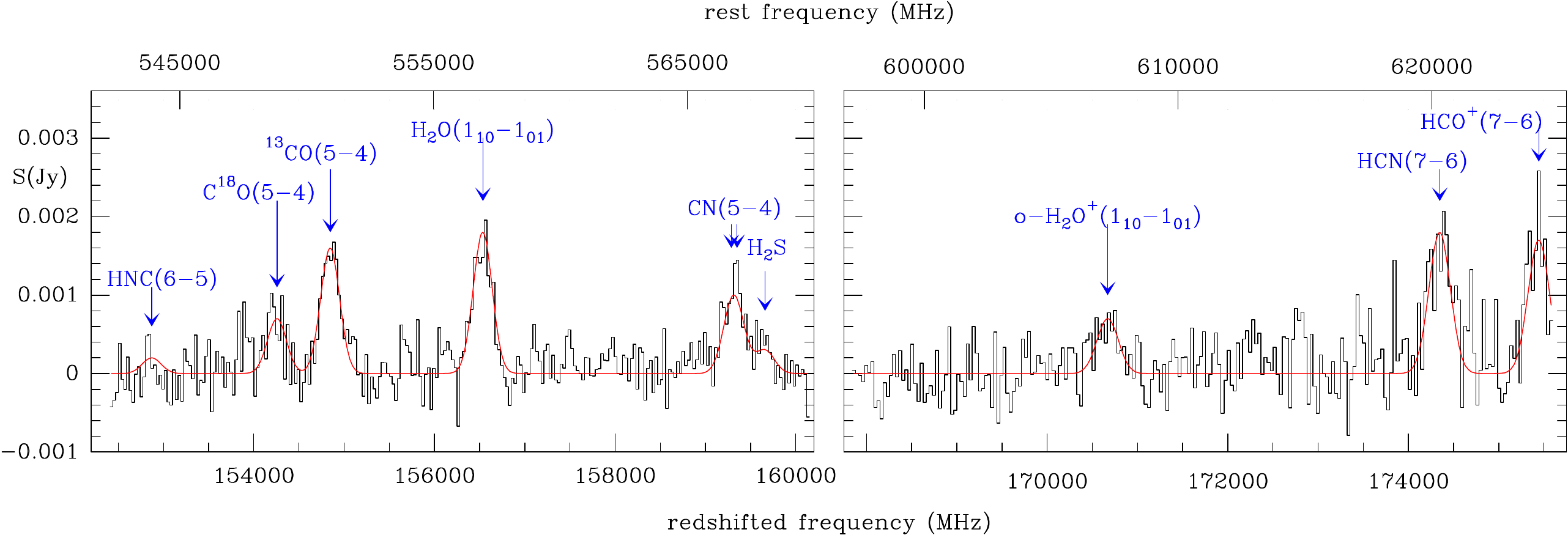}
\caption{NOEMA 2-mm spectrum of the Cloverleaf Quasar, covering lines
of H$_2$O, H$_2$O$^+$, H$_2$S, CN, HCN, HNC and HCO$^+$, as well as of the rare isotopologues
C$^{18}$O and $^{13}$CO. The 16 GHz-wide spectrum was observed with
one single frequency setting in 2 polarizations. {A flat (degree 0)
baseline has been subtracted from each 8 GHz-wide sideband. On-source
integration time is 8h} -- Gu\'elin et al. {\it in prep.} 
}\label{Clover-2mm}
\end{center}
\end{figure}

The molecular emission is currently the object of high sensitivity line surveys 
{using NOEMA and ALMA}.
Thirty four molecular lines pertaining to fourteen molecular species
are identified within the (still partly) surveyed 200 GHz-wide frequency band 
{covering the rest frequency ranges: 338 to 365 GHz, 526 to 626 GHz and 705 to 786 GHz.}   
(see Figs. \ref{Clover-3mm} and \ref{Clover-2mm}).  Some of those lines correspond to different rotational
transitions of the same species, such as the J$\rightarrow$J' = 1-0,
4-3, 6-5, 7-6, and 8-7 transitions of HCN and HCO$^+$, the 4-3, 6-5,
8-7 transitions of HNC, the 1-0, 3-2, 4-3, 6-5, 8-7, 9-8 transitions
of $^{12}$CO, the main isotopologue of CO, as well as the 5-4 and 7-6
transitions of the rare isotopologues $^{13}$CO and C$^{18}$O.

The detected lines sample a wide range of energy levels and 
opacities for a given molecule, and different excitation conditions for molecules that 
are believed to be coeval. From this, and from the spectral energy distribution of
the quasar continuum emission, it is possible to assess the density, temperature and radiation
field in the emitting region, as well as the molecular abundances. The constancy of the line width
(half-power width 440 kms$^{-1}$) seems to imply that the molecules are distributed throughout the
same rotating cloud complex.

{Besides the AGN emission, the far infrared} radiation field is mainly thermal emission from tiny dust grains
with temperatures in the range 30 K to 100 K, i.e. warm compared to
the TMC-1 Galactic cloud, but similar to that in the HII region
SgrB2 {(Weiss et al. 2003)}. Its intensity is however far more intense than in any Galactic
source, implying an exceptionally strong starburst: the mass of very dense gas is estimated to 
be at least $10^{10}$ Solar masses and the star formation rate \.{M}=$10^3$ Solar Masses per year.
This is 200 times larger than the star forming rate in the entire Milky-Way and comparable to those
of the brightest ultra luminous infrared galaxies, or ULIRGs.
(Solomon et al. 2003, Robitaille \& Whitney 2010).
 
In contrast to the absorption spectra observed toward PKS1830-211, where 
the line intensities primarily depend on the background quasar continuum flux, the  
lines detected in emission in a remote radio-quiet object like the Cloverleaf depend 
on its distance and are therefore very weak. No wonder that
most of the molecules observed there are di- or triatomic species. Some are
radicals and ions (CH, CCH, H$_2$O$^+$) characteristic of gas
irradiated by stellar UV, i.e. of diffuse clouds. The others (CO, CS,
HCN, HNC, HCO$^+$ H$_2$O, H$_2$S) are stable species belonging to
UV-shielded clouds. Remarkably, these stable species are the same as
the very first identified in Galactic clouds 50 years ago, when
the sensitivity of mm-wave radio telescopes precluded to detect more
than a dozen of molecular lines.

As much as we can tell {from the line intensities in Figs. \ref{Clover-3mm} and \ref{Clover-2mm}}, 
the abundances of HCN and HCO$^+$, relative to
the abundance of CO (the best tracer of the molecular gas), are similar in the 
Cloverleaf Quasar ($1-3\times 10^{-3}$) to those in the z=0.89 spiral galaxy and in the nearby 
Galactic cloud TMC-1 (see Muller et al. 2011), hence in an extremely wide range of environments and at all
Cosmic ages greater than 1 Gyr. {We note that the $^{13}$CO/C$^{18}$O isotopic abundance ratio on Fig. 
\ref{Clover-2mm}, 2.5, is typical of nearby star-forming galaxies and significantly larger than the values 
$\simeq 1$ reported by Zhang et al. (2018) for dust-enshrouded starbursts at similar redshift.}  

Water vapour is detected {and fairly abundant} in all those objects. It seems ubiquitous in remote 
quasars (Yang et al. 2016, {2019}), as it is in all Milky-Way molecular sources, where 
it is observed in both forms, gaseous and solid. {The H$_2$O submillimetre rotational lines, are
typically the strongest after those of CO. The H$_2$O and CO lines have similar profiles, which 
seems to imply both molecular species are coeval.}  

\vspace{1.0cm}
\section{Conclusion and Prospects}

We do know, from the spectral analysis of stellar light and
meteorites, that the elemental composition of matter is similar
throughout the Universe, at least as concerns the most abundant
elements: H, He, O, C, N, Ne, Si, Mg, S, P, Fe... It basically follows
the nuclear binding energies and, in the case of peculiar stars,
depends on their mass, which governs nucleosynthesis.

Things, of course, are less straightforward as concerns the gas
molecular composition, in view of the extraordinary large range of
physical and enviromental conditions prevailing in IS space.  It was
thought, for a long time, that the harsh conditions (typically
densities below few hundreds atoms per cubic cm, gas temperature of
few tens of kelvin, stellar and cosmic-ray induced UV radiation)
preclude the survival of any molecule, but for a few ephemeral
diatomic species like the already identified CH, CH$^+$ and CN. That
view was overturned at the end of the 1960s         
when it was realized that shielding by dust grains and H atoms, and
self-shielding of H$_2$, N$_2$ and CO molecules, considerably reduce
the photodestruction of those and very many other molecular species in
IS clouds. And indeed, in the following years formaldehyde, H$_2$CO and
CO were not only detected, but observed in about every dark or bright IS nebula. In
the course of years, owing to remarkable gains in sensitivity and
angular resolution of the mm-wave and far-IR telescopes, the list of
molecules identified in interstellar and circumstellar clouds has
steadily increased to reach 256 species {(see McGuire et al. 2018 for 
a list of molecules identified until 2018 and {\it The Astrochymist}
for a monthly update of that list.)}  

Most remarkably, HCN and H$_2$O are found to be ubiquitous in clouds
shielded by dust from stellar UV radiation.  As we have seen, they are
observed up to the edges of the Universe. Quite sensitive to stellar
UV and cosmic-ray induced photodestruction, these species cannot
survive for a very long time even in the densest clouds, hence have to be
renewed on relatively short timescales in a variety of environments.  HCN is a
prerequisite brick for the formation of polymers via peptide bonds,
whereas liquid H$_2$O, a solvant of polar molecules, is essential in
shaping up complex molecular structures. Their omnipresence in the gas
or on dust grains opens the way to organic molecules: formaldehyde,
formamide, methanol, ethanol, urea ...

Along this line, the recent gains in sensitivity, allowed by the new generation radio telescopes, 
brought an unexpected result: the identification in the quiescent dark cloud
TMC-1 of fairly complex organic compounds, linear chains or
cycles, consisting of 6 or more C-atoms. Latest examples are benzyne,
(c-C$_6$H$_4$), indene (c-C$_9$H$_8$ --Fig. \ref{indene_spectra}) and cyanoethyleneallene
(H$_2$C$_2$CHC$_3$N; Shingledecker et al., 2021) that were discovered this year together with a score of new
hydrocarbons or cyanohydrocarbons (the attachment of a CN radical
increases the polarity of hydrocarbons, easing their radio detection). Yet, 
TMC-1 would seem to be the last place where complex
molecules would form: its relatively short age (few x 10$^5$ years)
coupled to a vanishingly low gas density (few x 10$^4$ H$_2$ molecules
per cm$^3$), extremely low gas temperature (T$_k \simeq 10$ K), and to a total
lack of internal or nearby energy sources, should have made molecule formation a
formidable challenge. 

\begin{figure}[h!]
\begin{center}
\includegraphics[width=17cm,angle=0]{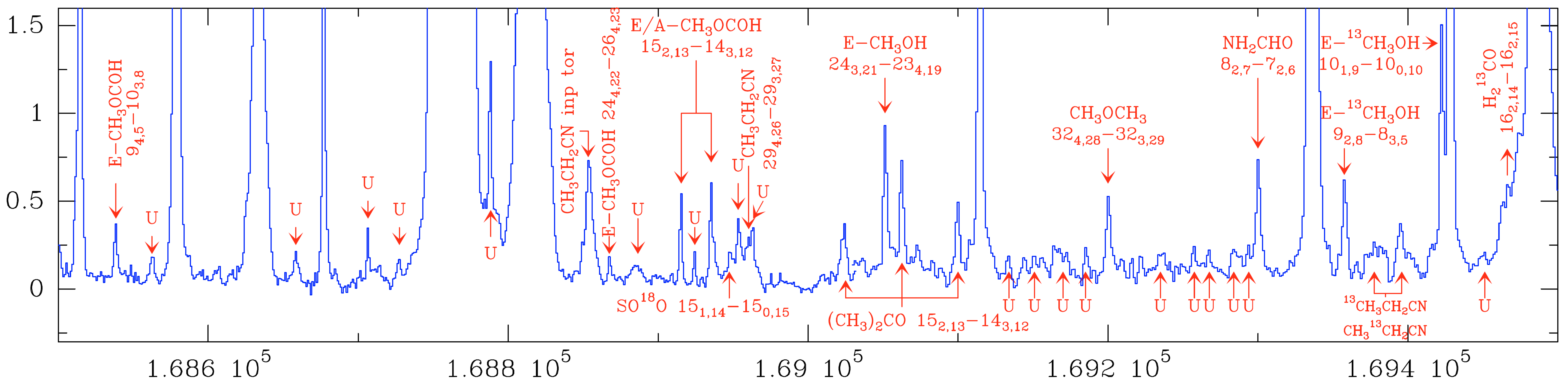} 
\caption{1 GHz-wide portion of the sub-millimeter spectrum of the Orion-KL 
star-forming cloud at 2mm. The spectrum, which is part of an extended (80 to 281 GHz) spectral 
survey made with the IRAM 30-m radio telescope, is confusion-limited above 200 GHz (--Tercero et al. 2010).}
\label{Orion-submm}
\end{center}
\end{figure}

The list of molecules identified in TMC-1 has more than doubled in the last two years, 
even though, due to the very low temperature,  most of the 
oxygen atoms are frozen on the dust grains in the form of water and dry ices, or trapped 
into CO molecules. For sure, many more will be discovered as low frequency searches for heavy 
organic molecules intensify. In warmer and denser sources, such as 
star-forming clouds and Giant Molecular clouds (OrionA-KL, SgrB2-N,...), not only reactions 
in the gas phase do proceed faster, but also the molecules formed on the grains may sublimate. 
Their molecular content is obviously richer than that of the cold dark clouds, especially as
concerns complex organic compounds. The census of those, however, becomes more
challenging, as the gas is more turbulent in such sources and the molecular lines broader, so that
the mm-wave (and far-IR) spectra reach much sooner the confusion limit (Fig. \ref{Orion-submm}). Prospects
are more favourable for Protoplanetary disks, which have narrow lines. These disks, 
however, are very small and are  barely resolved even with the largest interferometers. 
Their lines are very weak and, so far, only a score of molecular
species have been identified there. Those species, of course, include CO, HCN, HCO$^+$, H$_2$O, NH$_3$, H$_2$S, SO 
(Phuong et al. 2018), but also
formaldehyde, methyl cyanide, cyanoacetylene, formic acid and methanol. Many more will be detected with more sensitive 
searches in the near future.

\begin{figure}[!ht]
\begin{center}
\includegraphics[height=17cm,angle=-90]{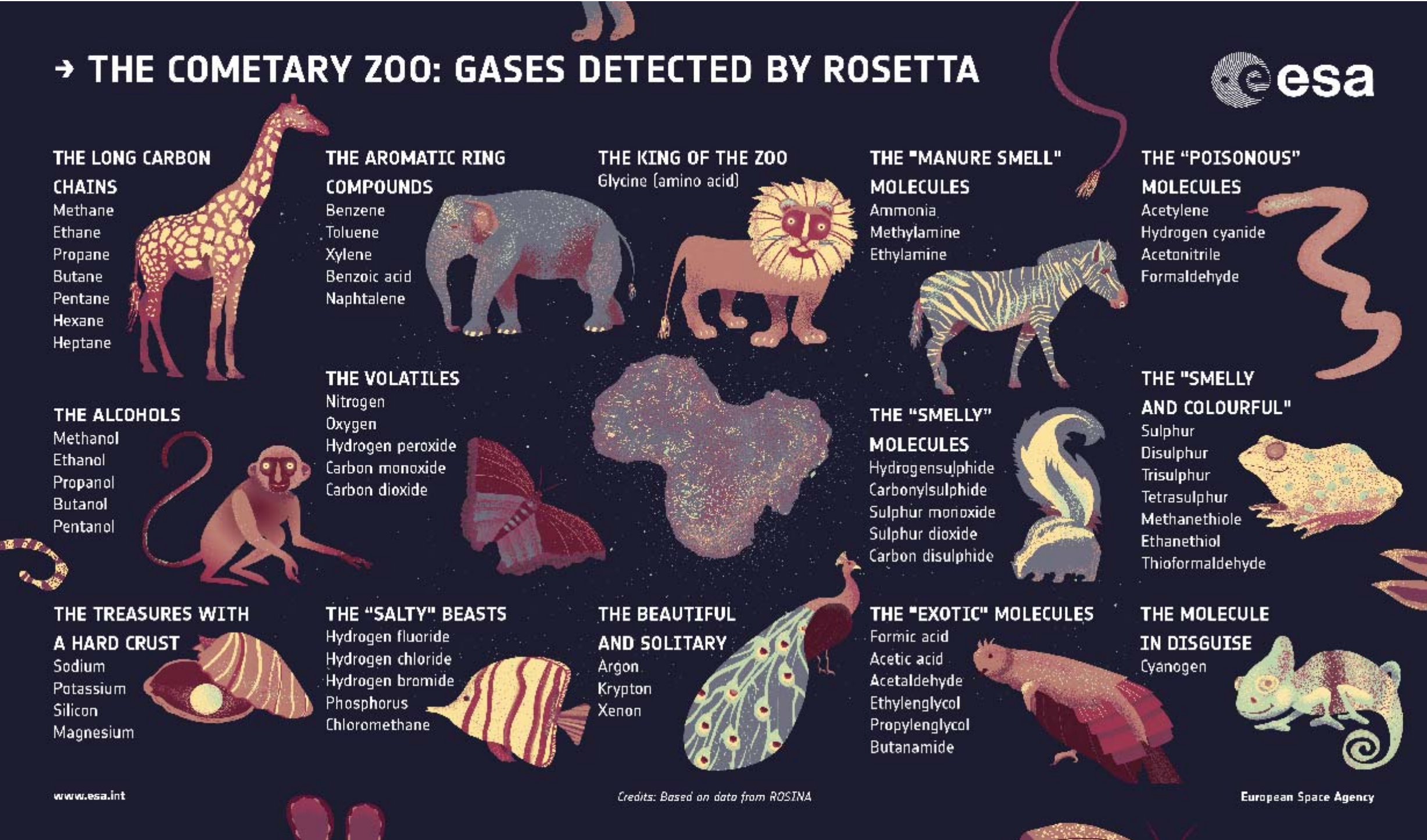}  
\caption{{List of gaseous atomic and molecular species detected by the Rosetta Orbiter Spectrometer (ROSINA) 
in the comet 67P/Churymov-Gerasimenko (Altwegg et al. 2019) here summarised in a humorous way as a 
{\it cometary zoo} by Kathrin Altwegg, ROSINA principal investigator.
}}
\label{Zoo}
\end{center}
\end{figure}

The bricks essential for the construction of amino acids and the
nucleobases seem therefore widespread across the Universe: 
didn't Huygens prophetically write in {\it The Cosmotheoros:
What's true in one part will hold over the whole Universe}?
Does this mean they were originally accreted by the Earth from outer
space? Not necessarily: it just shows those bricks form easily about
everywhere and may survive in harsh environmental conditions. Most
likely, they were already present in the pre-Solar nebula. They may
have been destroyed by the Sun's ignition and the Earth's formation, but
were quickly re-formed on the rocky/dusty surfaces of the young Earth
and of a number of Solar System bodies, in particular meteorites and
comets. The ``Cometary Zoo'' (Fig. \ref{Zoo}) observed on Comet
Churyumov-Gerasimenko by the Rosetta space mission, seems to
corroborate this view.

{The authors thank the referees for their constructive suggestions and
comments. J. Cernicharo thanks ERC for funding support through grant
ERC-2013-Syg-610256-NANOCOSMOS, and the Ministerio de Ciencia e
Innovaci\'on of Spain (MICIU) for funding support through projects
PID2019-106110GB-I00, PID2019-107115GB-C21 / AEI /
10.13039/501100011033, and PID2019-106235GB-I00.  This paper makes use
of ALMA data from project 2017.1.01232.S, of IRAM NOEMA interferometer
data from projects V03E, W19DA and S20BY, and of Yebes 40-m telescope
data from projects 19A003, 20A014, 20A017, 20D023, and 21A011.  ALMA
is a partnership of ESO (representing its member states), NSF (USA)
and NINS (Japan), together with NRC (Canada), MOST and ASIAA (Taiwan),
and KASI (Republic of Korea), in cooperation with the Republic of
Chile. IRAM is supported by INSU/CNRS (France), MPG (Germany) and IGN
(Spain). The Yebes 40-m telescope belongs to Instituto Geografico
Nacional (Spain). }

\subsection{references}
\noindent { Alloin, D., Guilloteau, S., Barvainis, R., Antonucci, R., and Tacconi, L. (1997).
The gravitational lensing nature of the Cloverleaf unveiled in CO (7-6) line emission.
A\&A, 321, 24}

\noindent { Altwegg, K., Balsiger, H. and Fuselier, S.A. (2019)
Cometary Chemistry and the Origin of Icy Solar System Bodies: The View after Rosetta.
ARA\&A, 57, 113}

\noindent { Ade, P.A.R., Aghanim, N., Arnaud, M., Baccigalupi, C., Banday, A.J. et al. (2016).
The Planck list of high-redshift source candidates.
A\&A, 596,A100}

\noindent Ag\'undez, M., Cernicharo, J., \& Gu\'elin, M. (2015a). Discovery of interstellar ketenyl (HCCO),
a surprisingly abundant radical. A\&A, 577, L5

\noindent Ag\'undez, M., Cernicharo, J., deVicente, P., Marcelino, N., Roueff, E., Fuente, A. et al. (2015b). 
Probing non-polar interstellar molecules through their protonated form: Detection of protonated 
cyanogen (NCCNH$^+$). A\&A, 579, L10

\noindent Ag\'undez, M., Marcelino, N., Tercero, B. et al. (2021a).
O-bearing complex organic molecules at the cyanopolyyne peak of TMC-1: 
Detection of C$_2$H$_3$CHO, C$_2$H$_3$OH, HCOOCH$_3$, and CH$_3$OCH$_3$. A\&A, 649, L4

\noindent Ag\'undez, M., Cabezas, C., Tercero, B., Marcelino, N., Gallego, J.D., DeVicente, P. et al. (2021b).
Discovery of the propargyl radical (CH$_2$CCH) in TMC-1: One of the most abundant radicals ever found and a key species for cyclization to 
benzene in cold dark clouds. A\&A, 647, L10

\noindent Allamandola, L.J., Tielens, A.G.G.M. \& Barker, J.R. (1985). 
Polycyclic aromatic hydrocarbons and the unidentified infrared emission bands: 
auto exhaust along the milky way. ApJ, 290, L25

\noindent Bañados, E.; Mazzucchelli, C.; Momjian, E.; Eilers, A-C; Wang, F; Schindler, J-T (2021) et al. (2021).
The Discovery of a Highly Accreting, Radio-loud Quasar at z = 6.82. ApJ 909, 80

\noindent Belloche, A., Menten, K.M., Comito, C., M\"uller, H.S.P., Schilke, P., Thorwirth, S. et al. (2008). 
Detection ofaminoacetonitrile in SgrB2(N). 
A\&A 482, 179

\noindent Bradford, M., J. E. Aguirre, Aikin, R., Bock, J.J., Earle, L., Glenn, J. et al. (2009).
The Warm Molecular Gas around the Cloverleaf Quasar.
ApJ, 705, 112
 
\noindent B$^2$FH: Burbidge, E. Margaret; Burbidge, G. R.; Fowler, William A.; Hoyle, F. (1957). 
Synthesis of the Elements in Stars. 
Rev. Modern Physics 29, 547

\noindent Burkhardt, A.M., Lee, L.K., Changala, B.P., Shingledecker, C. N., Cooke, I.R., Loomis, R.A., 
et al. (2021). 
Discovery of the Pure Polycyclic Aromatic Hydrocarbon Indene (c-C$_9$H$_8$) 
with GOTHAM Observations of TMC-1. 
ApJ, 913, L18

\noindent Cernicharo, J., \& Gu\'elin, M., Askne, J. (1984). 
TMC 1-like cloudlets in HCL 2.
A\&A, 138, 371

\noindent Cernicharo, J., \& Gu\'elin, M. (1987). The physical and chemical state of HCL2.
 A\&A, 176, 299

\noindent Cernicharo, J., Heras, A.M, Tielens, A.G.G.M., et al. (2001). 
Infrared Space Observatory's Discovery of C$_4$H$_2$, C$_6$H$_2$, and Benzene in CRL 618.
 ApJ, 546, L123

\noindent Cernicharo, J. (2004). 
The Polymerization of Acetylene, Hydrogen Cyanide, and Carbon Chains in the Neutral Layers of Carbon-rich Proto-planetary Nebulae.
ApJ, 608, L41

\noindent Cernicharo, J., Marcelino, N., Ag\'undez, M., Bermúdez, C., Cabezas, C., Tercero, B. et al. 2020a.
Discovery of HC$_4$NC in TMC-1: A study of the isomers of HC$_3$N, HC$_5$N, and H$_C7$N.
A\&A, 642, L17

\noindent Cernicharo, J., Marcelino, N., Pardo, J.R., Agúndez, M.; Tercero, B.; de Vicente, P. et al. (2020b). 
Interstellar nitrile anions: Detection of C$_3$N$^-$ and C$_5$N$^-$ in TMC-1. 
A\&A, 641, L9

\noindent Cernicharo, J., Marcelino, N., Ag\'undez, M., Endo, Y., Cabezas, C., Bermúdez, C. et al. (2020c).
Discovery of HC$_3$O$* +$ in space: The chemistry of O-bearing species in TMC-1.
A\&A, 642, L8

\noindent Cernicharo, J., Cabezas, C., Endo, Y., Marcelino, N., Agúndez, M., Tercero, B. et al. (2021a) 
Space and laboratory discovery of HC$_3$S$^+$.
A\&A, 646, L3

\noindent Cernicharo, J., Cabezas, C., Bailleux, S., Margulès, L., Motiyenko, R., Zou, L. et al. (2021b), 
Discovery of the acetyl cation, CH$_3$CO$^+$, in space and in the laboratory.
A\&A, 646, L7

\noindent Cernicharo, J., Ag\'undez, M., Kaiser, R.I., Cabezas, C.; Tercero, B.; Marcelino, N. et al. (2021c),
Discovery of benzyne, o-C$_6$H$_4$, in TMC-1 with the QUIJOTE line survey. 
A\&A, 652, L9

\noindent Cernicharo, J., Ag\'undez, M., Cabezas, C., Tercero, B., Marcelino, N., Pardo, J.R. et al. (2021d).
Pure hydrocarbon cycles in TMC-1: Discovery of ethynyl cyclopropenylidene, cyclopentadiene, and indene.
 A\&A, 649, L15

\noindent Cernicharo, J., Ag\'undez, M., Cabezas, C., Marcelino, N., Tercero, B., Pardo, J.R. et al. (2021e). 
Discovery of CH$_2$CHCCH and detection of HCCN, HC$_4$N, CH$_3$CH$_2$CN, and, tentatively, CH$_3$CH$_2$CCH in TMC-1.
A\&A, 647, L2

\noindent Cernicharo, J., Cabezas, C., Ag\'undez, M., Tercero, B., Marcelino, N., Pardo, J.R. et al. (2021f), 
Discovery of allenyl acetylene, H$_2$CCCHCCH, in TMC-1. A study of the isomers of C$_5$H$_4$.
A\&A, 647, L3

\noindent Cernicharo, J., Ag\'undez, M., Kaiser, R.I., Cabezas, C., Tercero, B., Marcelino, N., et al. (2021g), A\&A, in press

\noindent Cernicharo, J., Cabezas, C., Ag\'undez, M., Tercero, B., Pardo, J.R. Marcelino, N. et al. (2021h).
TMC-1, the starless core sulfur factory: Discovery of NCS, HCCS, H$_2$CCS, H$_2$CCCS, and C$_4$S and detection of C$_5$S.
A\&A, 648, L3

\noindent Cernicharo, J., Cabezas, C., Endo, Y., Ag\'undez, M., Tercero, B., Pardo, J.R.  et al. 2021i, 
The sulphur saga in TMC-1: Discovery of HCSCN and HCSCCH
A\&A, 650, L14

\noindent Cernicharo, J., Ag\'undez, M., Kaiser, R.I., Cabezas, C., Tercero, B. et al. 2021j, 
Discovery of two isomers of ethynyl cyclopentadiene in TMC-1: Abundances of CCH and CN derivatives of hydrocarbon cycles
A\&A, 655, L1

\noindent Cernicharo, J., Ag\'undez, M., Cabezas, C., Tercero, B., Marcelino, N. et al. 2021k, 
Discovery of HCCCO and C$_5$O in TMC-1 with the QUIJOTE line survey, A\&A, in press, arXiv:2112.01130

\noindent Cherchneff, I., Barker, J. R., and Tielens, A. G. G. M. 1992, 
Polycyclic Aromatic Hydrocarbon Formation in Carbon-rich Stellar Envelopes.
ApJ, 401, 269

\noindent Cheung, A.C., Rank, D.M., Townes, C.H., Thornton, D. D. and Welch, W.J. (1968).
Detection of NH3 Molecules in the Interstellar Medium by Their Microwave Emission.
PhysRevLett, 21, 1701

\noindent Cheung, A.C., Rank, D.M., Townes, C.H., Thornton, D. D. and Welch, W.J. (1968)
Detection of Water in Interstellar Regions by its Microwave Radiation.
Nature, 221, 626

\noindent Combes, F., Gupta, N., Muller, S., Balashev, S., Józsa, G.I.G., Srianand, R. et al. (2021).
PKS1830-211: OH and Hi at z = 0:89 and the first MeerKAT UHF spectrum.
A\&A, 648, 116

\noindent Cordiner, M. A., Charnley, S. B., Kisiel, Z., et al. (2017). 
Deep K-band Observations of TMC-1 with the Green Bank Telescope: Detection of HC$_7$O, Nondetection of HC$_{11}$N, and a Search for New Organic Molecules.
ApJ, 850, 187

\noindent Douglas, A.E. and Herzberg, G. (1941).
Note on CH$^+$ in Interstellar Space and in the Laboratory.
ApJ 94, 3

\noindent Ferus, M., Laitl, V., Knizek, A., Kubelík, P., Sponer, J., K\'ara, J. (2018).
HNCO-based synthesis of formamide in planetary atmospheres. 
A\&A 616, 150

\noindent Fuente, A., Navarro, D.G., Caselli, P., Gerin, M., Kramer, C., Roueff, E. et al. (2019).
Gas phase Elemental abundances in Molecular cloudS (GEMS). I. The prototypical dark cloud TMC-1.
 A\&A, 624, A105

\noindent {García, R. G.,  Gentaz O., Baldino, M. and Torres, M. (2012).
An 8 GHz digital spectrometer for millimeter-wave astronomy.
Millimeter, Submillimeter, and Far-Infrared Detectors and Instrumentation for Astronomy VI. 
Proceedings of the SPIE, Volume 8452, article id. 84522T} 

\noindent {García-Burillo, S., Graci\'a-Carpio, J., Guélin, M., Neri, R., Cox, P., Planesas, P. (2006) 
A New Probe of Dense Gas at High Redshift: Detection of HCO+ (5-4) Line Emission in APM 08279+5255 
ApJ, 645, L17}

\noindent {Green, S., Montgomery, J.A. and Thaddeus, P. (1974).
Tentative Identification of U93.174 as the Molecular Ion N$_2$H$^+$.
ApJ 193 L89}

\noindent { Guelin, M.; Thaddeus, P. (1977).
Tentative Detection of the C3N Radical.
ApJ, 212, L81}   

\noindent {Gu\'elin, M., Green,S.and Thaddeus, P. (1978).
Detection of the C4H radical toward IRC +10216. 
ApJ, 224, L27 1978}

\noindent  Gu\'elin, M. and Cernicharo, J. (1989).
Molecular Abundances in the Dense Interstellar and Circumstellar Clouds.
The Physics and Chemistry of Interstellar Molecular Clouds, Proceedings of the Symposium, Zermatt, Switzerland, Sept. 22-25, 1988. Lecture Notes in Physics, Vol. 331, edited by G. Winnewisser and J. T. Armstrong. 
Springer-Verlag, Berlin, 1989., p.337

\noindent Gu\'elin, M.; Salomé, P.; Neri, R.; García-Burillo, S.; Graci\'a-Carpio, J.; Cernicharo, J (2007)
Detection of HNC and tentative detection of CN at z = 3.9.
A\&A, 462, L45

\noindent Gu\'elin, M., Omont, A., Kramer, C.. Cernicharo, J., Tercero, B. Yang, C. et al. (2021) 
{\it in preparation}

\noindent 
{Guilloteau, S., Delannoy, J., Downes, D., Greve, A., Gu\'elin, M., Lucas, R. et al. (1992)
The IRAM interferometer on Plateau de Bure. A\&A, 262, 6}

\noindent {Herbst, Eric; Klemperer, W. (1974).
Is X-Ogen HCO+? ApJ 188 255}

\noindent {Hodge, J.A. and da Cunha, E. (2021)
High-redshift star formation in the ALMA era.
arXiv:2004.00934 {\it submitted for publication in Royal Soc. Open Science}} 

\noindent Holtom, P. D., Bennett, C. J., Osamura, Y., Mason, N. J., Kaiser, R. I. (2005).
A Combined Experimental and Theoretical Study on the Formation of the Amino Acid Glycine (NH$_2$CH$_2$COOH) 
and Its Isomer (CH$_3$NHCOOH) in Extraterrestrial Ices.
Ap.J. 626, 940

\noindent Hoyle, F. (1957).  The Black Cloud. {\it Penguin Modern Classics}

\noindent {Huygens, Christiaan (1698). Cosmotheoros, Book I {\it Adriaan Moetjens, The Hague}}

\noindent {Jimenez-Serra, I, Martin-Pintado, J., Rivilla, V.M., Rodriguez-Almeida, L., Alonso Alonso E. R., Zeng. S. et al. (2020).
Toward the RNA-World in the Interstellar Medium—Detection of Urea and Search of 2-Amino-oxazole and Simple Sugars.
AsBio, 20.1048J.} 

\noindent Joblin, C. and Cernicharo, J. (2018). 
Detecting the building blocks of aromatics
Science, 359, 156

\noindent {Kaifu, N.,Ohishi, M., Kawaguchi, K., Saito, S., Yamamoto, S., Miyaji, T. (2004).
A 8,8--50 GHz Complete Spectral Line Survey toward TMC-1.
PASJ, 56, 69} 

\noindent {Kneib, J.P., Alloin, D. and Pello, R. (1998).
Unveiling the nature of the Cloverleaf lens-system: HST/NICMOS-2 observations.
A\&A, 339, L65}

\noindent Lee, K. L. L., Changala, P. B., Loomis, R. A., Burkhardt, A. M., Xue, Ci, Cordiner, M. A. et al. 2021, 
Interstellar Detection of 2-cyanocyclopentadiene, C$_5$H$_5$CN, a Second Five-membered Ring toward TMC-1.
ApJ, 910, L2

\noindent L\'eger, A., Puget, J.L. (1984).
Identification of the "unidentified" IR emission features of interstellar dust?
A\&A, 137, L5

\noindent Lidman, C., Courbin, F., Meylan, G., Broadhurst, T., Frye, B. and  Welch, W. J. W. (1999).
The Redshift of the Gravitationally Lensed Radio Source PKS 1830-211.
ApJ 514, L57

\noindent {Martin, S., Mauersberger, R., Martin-Pintado, J. and Garcia-Burisso, S. (2006).
A 2 millimeter spectral survey of the starburst galaxy NGC 253.
ApJS, 164, 450}

\noindent McCarthy, M. C., Lee, K. L. K., Loomis, R. A., Burkhardt, A. M., Shingledecker, C. N., Charnley, S. B. et al. (2021). 
Interstellar detection of the highly polar five-membered ring cyanocyclopentadiene.
Nat. Astron., 5, 176

\noindent Marcelino, N., Cernicharo, J., Ag\'undez, M., Roueff, E.; Gerin, M.; Martín-Pintado, J. et al. (2007). 
Discovery of Interstellar Propylene (CH$_2$CHCH$_3$): Missing Links in Interstellar Gas-Phase Chemistry.
ApJ, 665, L127

\noindent Mart\'inez, L., Santoro, G., Merino, P.,  Accolla, M., Lauwaet, K., Sobrado, J. et al. (2020). 
Prevalence of non-aromatic carbonaceous molecules in the inner regions of circumstellar envelopes.
Nature Astron., 4, 97

\noindent Marcelino, N., Ag\'undez, M., Tercero, B.,  Cabezas, C., Bermúdez, C., Gallego, J. D. et al. (2020). 
Tentative detection of HC$_5$NH$^+$ in TMC-1.
A\&A, 643, L6

\noindent Matthews, H.E., Irvine, E., Friberg, F.M., Brown, R. D. and Godfrey, P. D. (1984).
A new interstellar molecule: tricarbon monoxide.
Nature, 310, 125

\noindent McGuire, B.A., Burkhardt, M., Shingledecker, C.N., Kalenskii, Sergei V., Herbst, E., Remijan, A. J. et al. (2017). 
Detection of Interstellar HC$_5$O in TMC-1 with the Green Bank Telescope.
ApJ, 843, L28

\noindent  {McGuire, B. A., Burkhardt, A. M., Kalenskii, S.; Shingledecker, C. N., Remijan, A. J., Herbst, E. and
 McCarthy, M. C. (2018a).
Detection of the aromatic molecule benzonitrile (c-C$_6$H$_5$CN) in the interstellar medium.
Science 359, 202}

\noindent {McGuire, B.A. (2018b).
Census of Interstellar, Circumstellar, Extragalactic, Protoplanetary Disk, and Exoplanetary Molecules
ApJS, 239, 1}

\noindent McGuire, B. A., Loomis, R. A., Burkhardt, A. M., Lee, K. L. K.; Shingledecker, C. N., Charnley, S. B. et al. (2021). 
Detection of two interstellar polycyclic aromatic hydrocarbons via spectral matched filtering.
Science, 371, 1265

\noindent {McKellar, A. (1940).
Evidence for the Molecular Origin of Some Hitherto Unidentified Interstellar Lines.
PASP 52, 187}

\noindent {Muller, S.\& Gu\'elin, M. (2008) 
Drastic changes in the molecular absorption at redshift z = 0:89 toward the quasar PKS 1830-211.
A\&A, 491, 739}

\noindent Muller, S.; Gu\'elin, M.; Dumke, M.; Lucas, R.; Combes, F. (2006).
Probing isotopic ratios at z = 0.89: molecular line absorption in front of the quasar PKS 1830-211. 
A\&A 458, 417

\noindent Muller, S.; Beelen, A.; Gu\'elin, M., Aalto, S., Black, J. H., Combes, F. et al. (2011).
Molecules at z = 0.89. A 4-mm-rest-frame absorption-line survey toward PKS 1830-211. 
A\&A 535, A.103  

\noindent Muller, S., Combes, F., Gu\'elin, M., G\'erin, M., Aalto, S., Beelen, A. et al. (2014).
An ALMA Early Science survey of molecular absorption lines toward PKS 1830-211. Analysis of the absorption profiles. A\&A...566, A.112

\noindent Muller, S., Müller, H. S. P., Black, J. H., Beelen, A., Combes, F., Curran, S. et al. (2016).
OH$^+$ and H$_2$O$^+$ absorption towards PKS 1830-211. 
A\&A 595, A.128. 

\noindent Muller, S., Müller, H. S. P., Black, J. H., G\'erin, M., Combes, F., Curran, S. (2017),
Detection of CH+, SH+, and their 13C- and 34S-isotopologues toward PKS 1830-211.
A\&A 606, 109

\noindent Muller, S.; Roueff, E.; Black, J. H., G\'erin, M., Gu\'elin, M., Menten, K. et al. (2020). 
Detection of deuterated molecules, but not of lithium hydride, in the z = 0.89 absorber toward PKS 1830-211.  
A\&A 637 

\noindent Ohishi, M., Suzuki, H., Ishikawa, S.-I., Yamada, C., Kanamori, H., Irvine, W. M. et al. (1991). 
Detection of a New Carbon-Chain Molecule, CCO.
ApJ, 380, L39

\noindent A. Pensabene1; 2, R. Decarli2, E. Bañados3, B. Venemans3, F. Walter3; 4, F. Bertoldi5
An ALMA multi-line survey of the ISM in two quasar host–companion galaxy pairs at z $>$ 6.

\noindent Pilleri, P., Joblin, C., Boulanger, F. and Onaka, T. (2015). 
Mixed aliphatic and aromatic composition of evaporating very small grains in NGC 7023 revealed by 
the 3.4/3.3 $\mu$m ratio.
A\&A, 577, A16

\noindent Phuong, N. T., Chapillon, E., Majumdar, L., Dutrey, A., Guilloteau, S., Piétu, V. (2018).
First detection of H$_2$S in a protoplanetary disk. The dense GG Tauri A ring. 
A\&A 616, L5

\noindent {Reuter,C., Vieira, J. D., Spilker, J. S., Weiss, A., Aravena, M., Archipley, M. et al. (2020)
The Complete Redshift Distribution of Dusty Star-forming Galaxies from the SPT-SZ Survey.
ApJ, 902, 78}

\noindent Rivilla, V. M., Martín-Pintado, J., Jiménez-Serra, I., Zeng, S., Martín, S., Armijos-Abendaño, J. et al. (2019).
Abundant Z-cyanomethanimine in the interstellar medium: paving the way to the synthesis of adenine.
MNRAS 483, L.114

\noindent {Robintaille, T.P. and Whitney, B.A. (2010).
The present-day star formation rate of the Milky-Way determined from Spitzer detected young stellar objects.
ApJ, 710, L11}

\noindent  Rubin, R. H. ; Swenson, G. W., Jr. ; Benson, R. C. ; Tigelaar, H. L. ; Flygare, W. H. (1971)
Microwave Detection of Interstellar Formamide 
ApJ 169, L39

\noindent Shingledecker, C.N., Lee, K.L.K., Wandishin, J.T., Balucani, N., Burkhardt, A. M. ; Charnley, S.B.,
et al. (2021). Detection of interstellar H$_2$CCCHC$_3$N. A possible link between chains and 
rings in cold cores, A\&A, 652, L12

\noindent {Snyder, L. E., Lovas, F. J., Hollis, J. M., Friedel, D. N., Jewell, P. R., Remijan, A. et al. (2005)
A Rigorous Attempt to Verify Interstellar Glycine 
ApJS 108, 301}

\noindent {Spilker, J. S., Marrone, D. P., Aguirre, J. E., Aravena, M. , Ashby, M. L. N. , Béthermin, M. et a;. (2014).
The Rest-frame Submillimeter Spectrum of High-redshift, Dusty, Star-forming Galaxies.
ApJ, 785, 149} 

\noindent {Spilker, J. S., Marrone, D. P., Aravena, M. , Béthermin, M., Bothwell, M.S., Carlstrom, J.E. et al. (2016).
ALMA Imaging and Gravitational Lens Models of South Pole Telescope --Selected Dusty, Sta-Forming Galaxies at High Redshifts.
ApJ, 826, 112}

\noindent {Solomon, P., Vanden Bout, P., Carilli, C. and  Gu\'elin, M. (2003).
The essential signature of a massive starburst in a distant quasar.
Nature, 426, 636}

\noindent {Stacey, H.R., McKean, J.P., Powell, D.M., Vegetti, S., Rizzo, F.,Spingola, C. (2021).
The rocky road to quiescence: compaction and quenching of quasar host galaxies.
MNRAS , 500, 3667}

\noindent {Swings, P. \& Rosenfeld, L. (1937)
Considerations Regarding Interstellar Molecules.
ApJ 86, 4}

\noindent Tercero, B., Cernicharo, J., Pardo, J.R., and Goicoechea, J. 2010
A line confusion limited millimeter survey of Orion KL. I. Sulfur carbon chains.
A\&A 517, A96

\noindent Tercero, B., Cernicharo, J., Cuadrado, S., de Vicente, P., Gu\'elin, M. (2020). 
New molecular species at redshift z = 0.89.
A\&A 636, L7

\noindent Tercero, F.; L\'opez-P\'erez, J. A.; Gallego, J. D.; Beltr\'an, F.; Garc\'ia, O.; Patino-Esteban, M. et al. (2021).
Yebes 40 m radio telescope and the broad band Nanocosmos receivers at 7 mm and 3 mm for line surveys.
A\&A 645, A..37

\noindent {Thaddeus, P. (2006).
The prebiotic molecules observed in the interstellar gas.
Phil. Trans. R. Soc. B 361, 1681}

\noindent {Tucker, K. D. ; Kutner, M. L. ; Thaddeus, P. (1974).
The Ethynyl Radical C2H-A New Interstellar Molecule.
ApJ 193, L115}

\noindent The Astrochymist {\it http://www.astrochymist.org/} 

\noindent {Venturini, S. and Solomon, P.L. (2003)
The Molecular Disk in the Cloverleaf Quasar.
ApJ, 590, 740}

\noindent Voltaire (1752) Micromegas Libretti Le Livre de Poche

\noindent {Wang, F., Yang, J., Fan, X., Hennawi, J.F., Barth, A.J., Banados, E. et al. (2021). 
A Luminous Quasar at Redshift 7.642.
ApJ, 907, L1}

\noindent {Weiß, A., De Breuck,C., Marrone, D.P., Vieira, J.D., Aguirre, J.E. ; Aird, K.A. ; Aravena, M. (2013)
 ALMA Redshifts of Millimeter-selected Galaxies from the SPT Survey: The Redshift Distribution of Dusty Star-forming Galaxies.
ApJ, 767, 88}

\noindent {Wildy, C. and Czerny, B. (2017).
The relationship between Mg II broad emission and quasar inclination angle.
FrASS, 4, 43}

\noindent {Wilson, R.W., Jefferts,K.B. and Penzias, A.A. (1970)
Carbon Monoxide in the Orion Nebula.   
ApJ 161 L43}

\noindent Winn, J., Kochanek, C., McLeod, B., Falco, E., Impey, C.D., Rix, H-W (2002). PKS 1830-211: A Face-on Spiral Galaxy Lens. 
ApJ...575..103

\noindent Woods, P.M., Millar, T. J., \& Zijlstra, A.A. and Herbst, E. (2002). 
The Synthesis of Benzene in the Proto-planetary Nebula CRL 618.
ApJ, 574, L167

\noindent Yang, C., Omont, A., Beelen, A., Gonz\'alez-Alfonso, E., Neri, R., Gao, Y. et al. (2016)  
   H$_2$O and H$_2$O$^+$ emission in lensed Hy/ULIRGs at z=2-4.
A\&A 595, A8 

\noindent {Yang, C., Gavazzi, A., Beelen, A. (2019).
CO, H$_2$O, H$_2$O$^+$ line and dust emission in a z=3.63 strongly lensed starburst merger at sun-kiloparsec scales.
A\&A, 624, 138.}

\noindent Yang, C., Gonz\'alez-Alfonso, E., Omont, A., Pereira-Santaella, M., Fischer, J., Beelen, A., 
  Gavazzi, R.et al. (2020). First detection of the 448 GHz ortho-H$_2$O line at high redshift. A\&A 634 L3

\noindent {Zhang, Z.Y., Romano, D., Ivinson, R.J., Papadopoulos, P.P.and Matteucci, F. (2018).
Stellar populations dominated by massive stars in dusty starburst galaxies across cosmic time.
Nature, 558, 260}

\noindent Zeng, S., Quénard, D., Jim\'enez-Serra, I., Mart\'in-Pintado, J., Rivilla, V. M., Testi, L., Martín-Doménech, R. (2019).
First detection of the pre-biotic molecule glycolonitrile (HOCH$_2$CN) in the interstellar medium.
MNRAS 484, L43

\subsection{figures}

\begin{figure}[h]
\begin{center}
\includegraphics[height=7cm,angle=0]{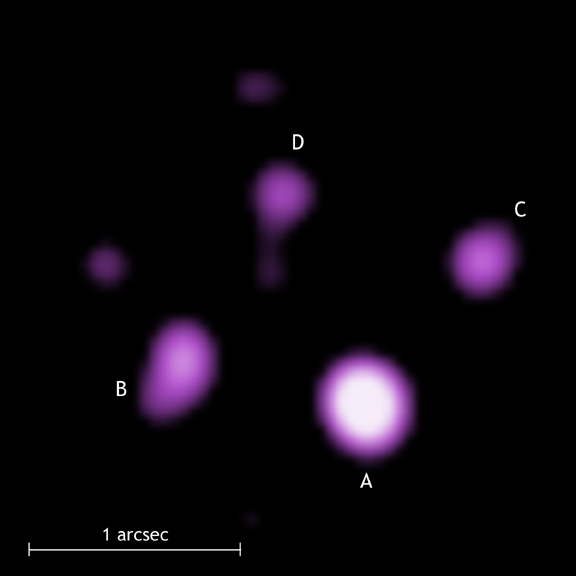}
\caption{The Cloverleaf Quasar, Chandra Xray image. Credit: NASA}\label{Chandra} 
\end{center}
\end{figure}

\vspace{-10cm}     

\end{document}